\documentclass[%
reprint,
showpacs,
bibnotes,
amsmath,amssymb,
prl,
floatfix,
]{revtex4-1} 

\usepackage{graphicx}
\usepackage{dcolumn}
\usepackage{bm}
\usepackage{hyperref}
\usepackage{multirow}
\usepackage{array}
\usepackage{booktabs}
\usepackage{ctable}
\usepackage{upgreek}
\usepackage{mathrsfs}
\usepackage{amssymb}
\usepackage{amsbsy}
\usepackage{color}
\usepackage{cancel}
\usepackage[bbgreekl]{mathbbol}
\usepackage{marginnote}
\usepackage{amsfonts}
\usepackage[caption=false]{subfig}
\newcommand{\bcen}{\begin{center}}
\newcommand{\ecen}{\end{center}}
\newcommand{\btab}{\begin{tabular}}
\newcommand{\etab}{\end{tabular}}
\newcommand{\bdes}{\begin{description}}
\newcommand{\edes}{\end{description}}

\newcommand{\beq}{\begin{equation}}
\newcommand{\eeq}{\end{equation}}
\newcommand{\bea}{\begin{eqnarray}}
\newcommand{\eea}{\end{eqnarray}}

\newcommand{\half}{\frac{1}{2}}
\newcommand{\bary}{\begin{array}}
\newcommand{\eary}{\end{array}}
\newcommand{\benum}{\begin{enumerate}}
\newcommand{\eenum}{\end{enumerate}}
\newcommand{\bitem}{\begin{itemize}}
\newcommand{\eitem}{\end{itemize}}

\newcommand{\mean}[1]{\langle #1 \rangle}

\newcommand{\eqn}[1] {eqn.~(\ref{#1})}

\newcommand{\fig}[1]{fig.~\ref{#1}}
\newcommand{\Fig}[1]{Fig.~\ref{#1}}

\makeatletter

\newcommand{\Rmnum}[1]{\expandafter\@slowromancap\romannumeral #1@}
\makeatother

\newlength{\myfigwidth}
\newlength{\myhalffigwidth}
\newcommand{\kl}{{k_\ell}}
\newcommand{\klg}[1]{\protect{{k_\ell^{#1}}}}

\newcommand{\dW}{{\mathbb{W}}}

\newcommand{\dU}{{\mathbb{U}}}

\newcommand{\da}{{\mathbb{a}}}

\newcommand{\db}{{\mathbb{b}}}
\newcommand{\dC}{{\mathbb{C}}}

\newcommand{\dS}{{\mathbb{S}}}
\newcommand{\dOmg}{{\mathbb{\Omega}}}
\newcommand{\domg}{{\mathbb{\bbomega}}}
\newcommand{\dPhi}{{\mathbb{\Phi}}}

\newcommand{\mylabel}[1]{\label{#1}} 

\newsavebox{\measurebox}

\usepackage{bbm}

\usepackage{lineno}

\begin{document}



\title{Baryon squishing in synthetic dimensions by effective $SU(M)$ gauge fields}
\author{Sudeep Kumar Ghosh}
\email{sudeep@physics.iisc.ernet.in}
\author{Umesh K. Yadav}
\email{umesh@physics.iisc.ernet.in}
\author{Vijay B. Shenoy}
\email{shenoy@physics.iisc.ernet.in}
\affiliation{Centre for Condensed Matter Theory, Department of Physics, Indian Institute of Science, Bangalore 560 012, India}


\date{\today}

\begin{abstract}
We investigate few body physics in a cold atomic system with synthetic dimensions (Celi et al., PRL 112, 043001 (2014)) which realizes a Hofstadter model with long-ranged interactions along the synthetic dimension. We show that the problem can be mapped to a system of particles (with $SU(M)$ symmetric interactions) which experience an $SU(M)$ Zeeman field at each lattice site {\em and} a non-Abelian $SU(M)$ gauge potential that affects their hopping from one site to another. This mapping brings out the possibility of generating {\em non-local} interactions (interaction between particles at different physical sites). It also shows that the non-Abelian gauge field, which induces a flavor-orbital coupling, mitigates the ``baryon breaking'' effects of the Zeeman field. For $M$ particles, the $SU(M)$ singlet baryon which is site localized, is ``deformed'' to be a nonlocal object (``squished'' baryon) by the combination of the Zeeman and the non-Abelian gauge potential, an effect that we conclusively demonstrate by analytical arguments and exact (numerical) diagonalization studies. These results not only promise a rich phase diagram in the many body setting, but also suggests possibility of using cold atom systems to address problems that are inconceivable in traditional condensed matter systems. As an example, we show that the system can be adapted to realize Hamiltonians akin to the $SU(M)$ random flux model.

\end{abstract}


\maketitle


Emulation of quantum systems of interest to a wide range of physics from condensed matter, high energy field theory etc.,~is made possible with cold atoms.\cite{Bloch2008}  Recent experimental developments in the field stand testimony to this.\cite{Lin2009B,Lin2009A,Wang2010,Cheuk2012,Katerly2013,Bloch2013,Jotzu14} Adding to the excitement and soaring interest in this area is the possibility of realizing systems that have novel physics and yet exceedingly difficult or not realizable in conventional condensed matter systems. Systems with $SU(M)$ ($M>2$) symmetries are one such example. Indeed, $SU(M)$ symmetric spin models have interesting phases and phase transitions, \cite{Read89,Harada03,Assaad05,Corboz11,Xu10,Paramekanti07,Hermele11,Greiter07} as do Hubbard models with $SU(M)$ symmetry.\cite{Buchta07,Manmana11,Cai13,Zhou14} Several theoretical\cite{Hofstetter02,Honerkamp04,Klingschat10,Rapp08,Capponi08,Rapp07,Pohlmann13,Grass14} and experimental\cite{Cazalilla09,Cazalilla14,DeSalvo10,Fukuhara07,Prevedelli99,Gor10,Taie12} works have  explored physics of cold atomic systems with $SU(M)$ symmetry. Some of experimentally realized $SU(M)$ systems are $^{6}Li$($M = 4$) 
\cite{Chin06}, $^{173}Yb$ ($M = 6$)\cite{Fukuhara07,Taie12,Mancini15}, 
and $^{87}Sr$($M = 10$).\cite{DeSalvo10,Gor10,Cazalilla14}

\citeauthor{Celi2014}\cite{Celi2014} proposed the concept of ``synthetic dimensions'' which achieves the goal of realizing finite sized ``strip'' of a Hofstadter  model. Their idea, illustrated in \Fig{fig:schematic_plot}, involves atoms with $M$ internal states (labeled $1 \ldots \gamma$) in a 1D (this can also be in higher dimensions) optical lattice. The hopping of the atoms from a site $j$ (with coordinate $x_j = j d$, $d$ is the spacing of the optical lattice) to its neighbour does not change its internal state, and the amplitude $t$ is independent of $\gamma$. The internal states at a site $j$ are now coherently coupled such that an atom in state $\gamma$ at site $j$ can ``hop'' to the state $\gamma+1$ at $j$ with an amplitude $\Omega_\gamma^{j}$. This produces, as shown in \Fig{fig:schematic_plot}, a square lattice strip of finite width with $M$ sites along the ``synthetic dimension''. Since the coherent coupling is produced by a light of wavenumber $k_\ell$, we have $\Omega_\gamma^{j} = \Omega_\gamma e^{-i k_\ell x_j}$, and this results in an atom picking up a phase factor $e^{-ik_\ell d}$ upon hopping around a plaquette. Choosing $k_\ell d = 2 \pi \frac{p}{q}$ where $p$ and $q$ are relative prime integers, provides a realization of a finite strip of the Hofstadter model\cite{Jain2007}  with a $p/q$ flux per plaquette. Very recent experimental realization\cite{Mancini15} of this scheme bolsters the possibilities and scope of this research direction.

Another interesting aspect of the problem is that the $SU(M)$ symmetric interactions between the atoms at a site $j$ manifest as ``infinite-ranged'' (distance-independent) interactions along the synthetic dimension. For example, two atoms at site $j$ (see \Fig{fig:schematic_plot}) with $\gamma=1$ and $\gamma=2$ will interact with the same strength as $\gamma=1$ and $\gamma =4 (M)$. It is the physics of such a system that is the subject of this paper, i.~e., to understand interplay between the flux $p/q$ and the $SU(M)$ interactions. It is essential to focus, as we do, on the physics of few particles since it provides crucial insights into constructing a many body phase diagram of the system. Previous studies\cite{Rapp07,Capponi08,Klingschat10,Pohlmann13} of fermionic atoms with attractive $SU(M)$-interactions in a simple $1$d  lattice (no flux, i.~e., $\frac{p}{q} = 0, \Omega_\gamma=0$)  show the existence of $SU(M)$ singlet ``baryons'' and quasi-long-range color superfluidity of these baryons (see \cite{Guan2013} for a review). The central question that we address in this paper is the fate of these baryons in the presence of flux in the synthetic dimension.

Here we show that the Hofstadter model with infinite-ranged interaction along the synthetic dimension\cite{Celi2014} can be mapped to a problem of $M$-flavor particles with $SU(M)$ symmetric interactions hopping on the physical lattice with an on-site $SU(M)$ Zeeman potential (determined by the couplings $\Omega_\gamma$) along with a $SU(M)$ gauge field (determined by the flux $p/q$, and $\Omega_\gamma$) that controls their hopping from site to site. Further analysis along these lines reveals, {\em inter alia}, a) the gauge field which induces a flavor-orbit coupling mitigates the ``baryon breaking'' effects of the Zeeman field, and b) the gauge field also induces a {\em non local} interaction, i.e., interaction between particles at different $j$ sites. One crucial outcome is that under favourable circumstances, the $SU(M)$ singlet baryon ($\Omega_\gamma=0$), which is an object localized at a site $j$ but extended along the synthetic dimension, is transformed into an $M$-body bound state that is extended in real space (along $j$) which we dub as the ``squished baryon''. This is demonstrated by analytical arguments supported by detailed few body exact diagonalization calculations. This work not only suggests novel many-body phases (eg, ``squished baryon'' condensate) of these systems, but also suggests new opportunities with $SU(M)$ symmetric systems. For example, our mapping of the system to a set of particles with $SU(M)$ gauge field brings out the possibility of using the $M$-component cold atom systems to simulate the $SU(M)$ random flux model.\cite{Altland1999}

\begin{figure}
{
\centering
\includegraphics[height=0.75\myfigwidth,width=1.0\myfigwidth,trim= 0 50 0 0]{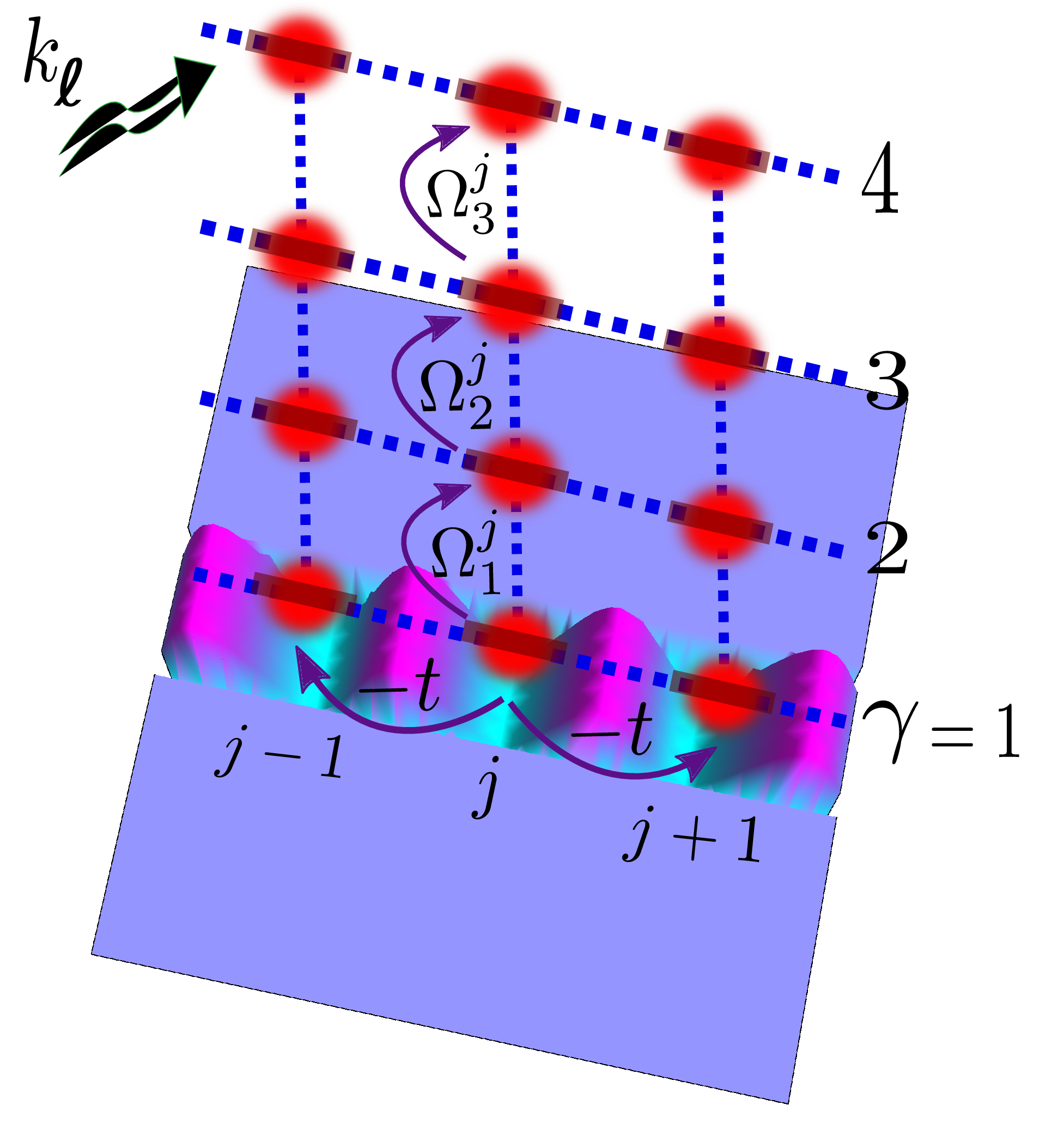}
}
\caption{(Color online) {\bf Synthetic dimension}: Atoms with $M$ ($M=4$ is illustrated) internal states (labelled by $\gamma$) hop on a 1D optical lattice (indicated by the alternating pink and cyan pattern) with amplitude $t$. The internal hyperfine states (indicated by thick brown lines with a red halo) are coupled coherently by a light of wave number $k_\ell$. The coupling at the $j$-th site between the state $\gamma$ and $\gamma +1 $ is denoted by $\Omega_\gamma^j$. The hyperfine states serve as an additional spatial dimension -- the synthetic dimension. Since the phases of $\Omega^j_\gamma$ depend on $j$, a particle hopping around a plaquette of this synthetic lattice picks up a phase equivalent to having a magnetic flux in the plaquette.\mylabel{fig:schematic_plot}
}
\end{figure}

\noindent
{\bf Methodology:} Denoting the operators that creates a fermion\footnote{While we discuss fermionic physics in this paper, many of our conclusions will be applicable also to bosonic systems.} at site $j$ with a hyperfine flavor $\gamma$ as $C^\dagger_{j,\gamma}$, the Hamiltonian is  ${\cal H} = H_t + H_\Omega + H_U$, with 
\begin{align}
H_t & = -t \sum_{j,\gamma=1,M} \left(C^\dagger_{j+1,\gamma}C_{j,\gamma} + \mbox{h.~c.}\right) \mylabel{eqn:Ht} \\
H_\Omega &= \sum_{j,\gamma=1,M-1} \left(\Omega_\gamma^j C^\dagger_{j,\gamma+1} C_{j,\gamma} + \mbox{h.~c.} \right) \mylabel{eqn:HOmg} \\
H_U & = - \frac{U}{2} \sum_{j,\gamma\gamma'} C^\dagger_{j,\gamma} C^\dagger_{j, \gamma'} C_{j,\gamma'}C_{j,\gamma} \mylabel{eqn:HU}
\end{align}
where $t$ is the intersite hopping, $U$ is the strength of the attractive $SU(M)$ symmetric interaction. The couplings $\Omega_\gamma^j= \Omega_\gamma e^{-ik_\ell x_j}$, where $\Omega_\gamma$ depend on details (see \cite{Celi2014,Mancini15}) of the system. 

A mapping gains further insights into the physics of the problem. Towards this end we introduce the notation where $\dC_j $ is the column $(C_{j,1}, C_{j,2},\ldots,C_{j,M})^T$. We introduce a local unitary transformation $\dC_j = \dW_j \db_j$ where $\dW_j = \mbox{Diag}\{e^{-i \klg{\gamma}x_j}\}, \gamma=1,M$, with $\klg{\gamma} = (\gamma -1) \kl$, and $\db_j = (b_{j,1},\ldots,b_{j,M})^T$ is another set of fermionic operators. Immediately, $H_\Omega = \sum_j \db^\dagger_j \dOmg \db_j$ where $\dOmg$ is a site independent Hermitian matrix 
\beq
\dOmg = \left( \begin{array}{ccccc}
0 & \Omega_1^* & 0 & \ldots & 0 \\
\Omega_1 & 0 & \Omega_2^* & \ldots & 0 \\
0 & \Omega_2 & 0 & \ldots &  \vdots \\
\vdots & \vdots & \vdots & \ddots & \Omega^*_{M-1} \\
0 & 0 & \ldots & \Omega_{M-1} & 0
\end{array} \right) \,\,.
\eeq
Diagonalization gives $\dOmg = \dS \domg \dS^\dagger$ where $\domg = \mbox{Diag}\{\omega_\zeta\}$,  $\zeta=1, \ldots, M$ is the diagonal matrix with eigenvalues $\omega_{\zeta}$, and $\dS$ is a unitary matrix. This results in $H_\Omega = \sum_j \da^\dagger_j \domg \da_j$, where $\dS^\dagger \db_j = \da_j = \left\{a_{j,\zeta} \right\}^T$, $\zeta=1,\ldots,M$ is a new set of fermion operators. Clearly, $\dC_i = \dU_i \da_i$ where $\dU_i = \dW_j \dS$ is a unitary matrix. We now have
\beq\mylabel{eqn:GF}
H = -t \sum_j \left(\da^\dagger_{j+1} \dU^\dagger_{j+1} \dU_j \da_j + \mbox{h.~c.} \right) + \sum_j \da^\dagger_j \domg \da_j +  H_U
\eeq
where $H_U$ is operator defined in \eqn{eqn:HU} written in terms of $a_{j,\zeta}$ owing to its $SU(M)$ invariance. We immediately see that in terms of the transformed states $\da_j$, the Hamiltonian can be interpreted as that of particles in a flavour ($\zeta$) dependent potential $\omega_\zeta$ (which is a $SU(M)$ Zeeman field), and whose hopping is influenced by a non-Abelian gauge field $\dU^\dagger_{j+1} \dU_j=\dS^\dagger\dPhi\dS$, ($\dPhi = \mbox{Diag}\left\{e^{i \klg{\zeta} d} \right\}$) that produces flavor mixing up on hopping from site to site. The Zeeman field is determined solely by $\Omega_\gamma$, while the gauge field has a crucial additional dependence on the flux $p/q$. We have thus mapped the problem of synthetic dimensions to a system of $M$ component fermions experiencing $SU(M)$ Zeeman and gauge fields and $SU(M)$ symmetric interactions.

\begin{figure}
\centerline{\includegraphics[width=1.1\myfigwidth]{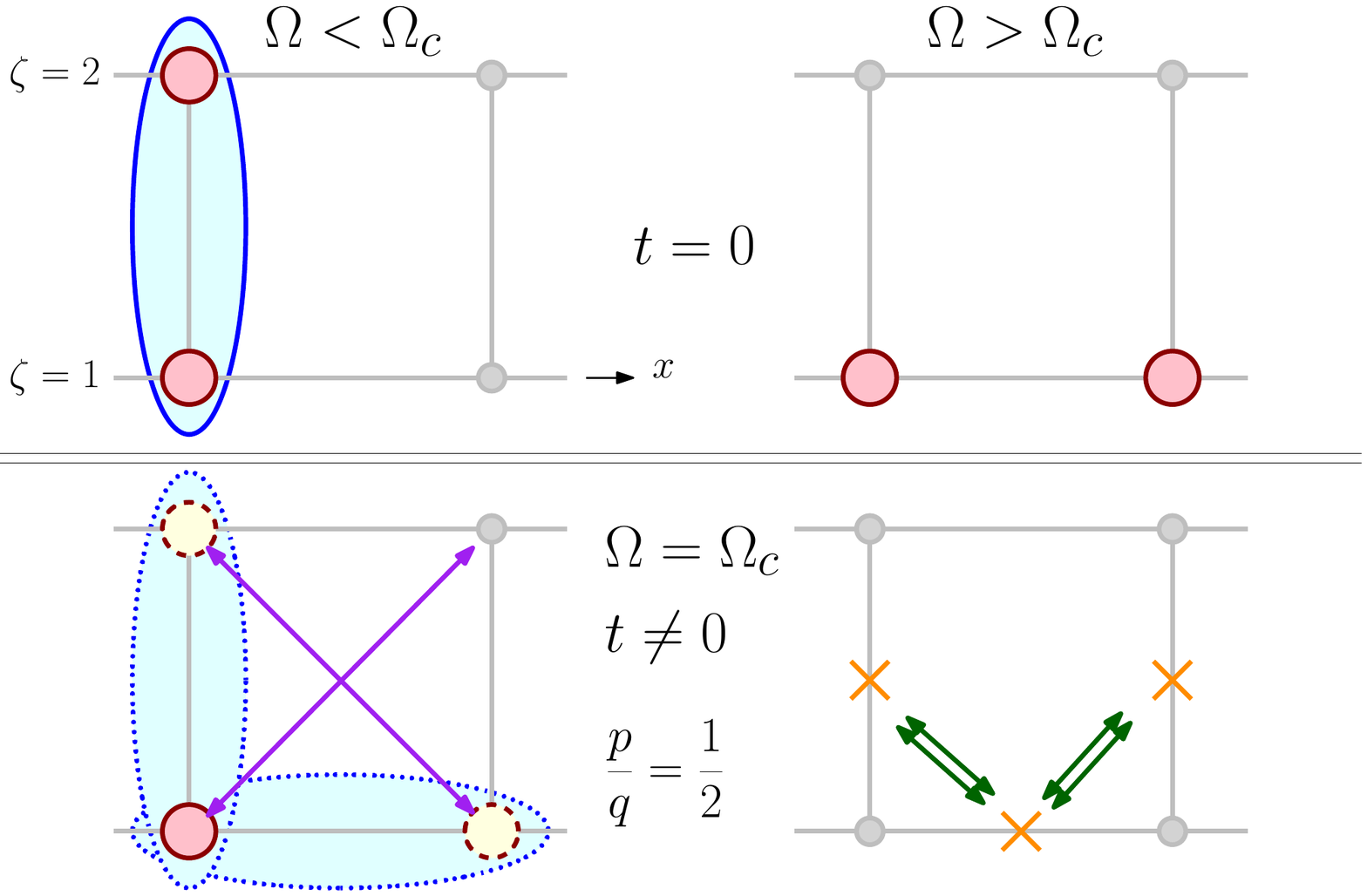}}
\caption{(Color online) {\bf Non-local interaction:} Top panel shows the state of two fermions when $t=0$ with $M=2$ which has $\omega_{\zeta=1}= - \Omega$ and $\omega_{\zeta = 2} = \Omega$. For $\Omega < \Omega_c$, $\Omega_c = \frac{U}{2}$, the usual baryon which is a bound state with both the $\zeta$ at a site being occupied (top left). When $\Omega > \Omega_c$, baryon breaking effect of the Zeeman field sets in, and the lowest energy state is each of the two particles occupying $\zeta=1$ state at arbitrary but distinct sites. At $\Omega=\Omega_c$, the two states are degenerate. Arrows in the left bottom panel show the hopping pattern when $t\ne0$ in the presence of a $\frac{1}{2}$-flux ($\frac{p}{q} = \frac{1}{2}$). If the two particles of the broken baryon are in the neighbouring sites as shown, then this baryon can effectively hop on a dual lattice shown by crosses (bottom right) by hybridizing with the degenerate baryon (vertical shaded bond), gaining kinetic energy. This produces a net attractive interaction between particles at neighbouring site with $\zeta=1$ and leads to ``baryon squishing'', i.~e., increasing non-local character of the bound state.
}
\mylabel{fig:illus}
\end{figure}

\noindent
{\bf Induced Interactions:} We now discuss the key outcome that arises in the analysis of \eqn{eqn:GF}.\footnote{We briefly mention that the mapping provides many insights that we do not explicitly discuss. For example, the exact solution of the two particle problem in the limit of $p/q \to 0$ can be obtained. In this limit, the flavor $\zeta$ is conserved, and for two particles with distinct flavours $\zeta$ and $\zeta'$, one obtains a ground state energy of $\omega_\zeta + \omega_{\zeta'}- \sqrt{U^2 + 16t^2}$}  Consider the $M=2$ system with $p/q=1/2$, i.e., the so called $\frac{1}{2}$-flux case. The main idea is uncovered by looking at the rather unnatural limit of vanishing hopping $t \to 0$. The Zeeman field in this case is $\domg = \mbox{Diag}\{-\Omega,\Omega\}$.  When $\Omega \ll U$, the ground state of the system is given by a $M=2$ baryon with two particles localized at the same site (\Fig{fig:illus}, top-left). The ``baryon breaking'' effect of the Zeeman field occurs when  $\Omega$ exceeds $\Omega_c=\frac{U}{2}$ (see \Fig{fig:illus}, top-right). In the broken baryon state, the particles both have $\zeta=1$ and can be located at any two sites with distinct $j$. Turning on $t$, $t > 0$ with a $\half$-flux now produces a hopping pattern of individual particles as shown by arrows in \fig{fig:illus} (bottom left). The hopping does not conserve the hyperfine flavour -- this flavour-orbital coupling is a result of the gauge potential that depends on the flux. It is clear that the degeneracy of the broken baryon states is lifted by the hopping with flavor-orbit coupling -- two particles with $\zeta=1$ the  on neighboring sites can gain energy by hybridizing with the degenerate baryon state (bound along the synthetic dimension). The effect just discussed, therefore, induces a {\em non local} attractive interaction between particles with $\zeta=1$ located on two neighbouring sites. Then net outcome of this is a ``squished baryon'' state that generically has a bound state character along the synthetic and real dimension. For the particular case just discussed, the state is a bound state of two particles that ``resonates'' between the vertical and horizontal bonds (as indicated bottom left figure of \fig{fig:illus}), hopping on the ``dual lattice'' indicated by crosses in \fig{fig:illus} (bottom right). Indeed, as $\Omega \gg \Omega_c$, the bound state is primarily made of particles with $\zeta=1$ -- ``fully'' squished baryon which is a result of the attractive interaction between near-neighbour $\zeta=1$ states which is proportional to $\frac{t^2}{2 \Omega - U}$.

Similar physics applies to a generic $M$.  An important point to be noted is that the scale $\Omega_c$ and the resulting ``broken baryon'' state depends on the details of the coupling $\Omega_\gamma$. For a given $M$ and $\Omega_\gamma$, our arguments suggests that there are some special fluxes $p/q$ that could more effectively produce the effect of the non-local binding and baryon squishing just discussed.
 
\noindent
{\bf Exact diagonalization:} We have numerically investigated the few body ($N_p$ particle) physics in this system using $N_q$ sites of the physical lattice with periodic boundary conditions, resulting in a Hilbert space size of $^{N_q M}C_{N_p}$. Our diagonalization scheme uses the translational symmetry, with $Q$, the total center of mass momentum, being the associated good quantum number. If overall ground state is found to occur with $Q=Q_g$, then the binding energy is defined as $E_b=E_g(Q_g,U=0)-E_g(Q_g,U)$, where $E_g(Q_g,U=0)$ is the ground state energy of the same system but without interactions between the particles. We also study the properties of ground states by computing the moment of inertia along the $x$-direction, and average value for the synthetic coordinate $\zeta$,
\begin{align}
\mylabel{eqn:Ixx} I_{xx} &= \frac{1}{^{N_p}C_{2}} \langle \sum_{i_1 > i_2} (\Delta x_{i_1,i_2})^2 \rangle \,\,,\\
\mylabel{eqn:zeta} \mean{\zeta} &= \frac{1}{N_p} \langle \sum_{i} \zeta_i \rangle\,\,,
\end{align}
where $i$-s run over the particle labels and $\Delta x_{i_1,i_2}= (x_{i_1}-x_{i_2})$.
An indicator of an $N_p$-particle  bound state is a positive binding energy $E_b>0$ and, $I_{xx}$ both of which are insensitive to $N_q$ (system size).\footnote{In a completely unbound state, such as that obtained with $U=0$, $I_{xx}\sim N_q^2$.} The quantity $\mean{\zeta}$ provides a measure of squishing. For example, with $N_p=M$, $\mean{\zeta} = \frac{(M+1)}{2}$ indicates the usual $SU(M)$-singlet baryon, while squishing can be deduced from a value of $\mean{\zeta} < \frac{(M+1)}{2}$.

\begin{figure*}
{
\centering
\includegraphics[height=0.8\myfigwidth,width=1.15\myfigwidth]{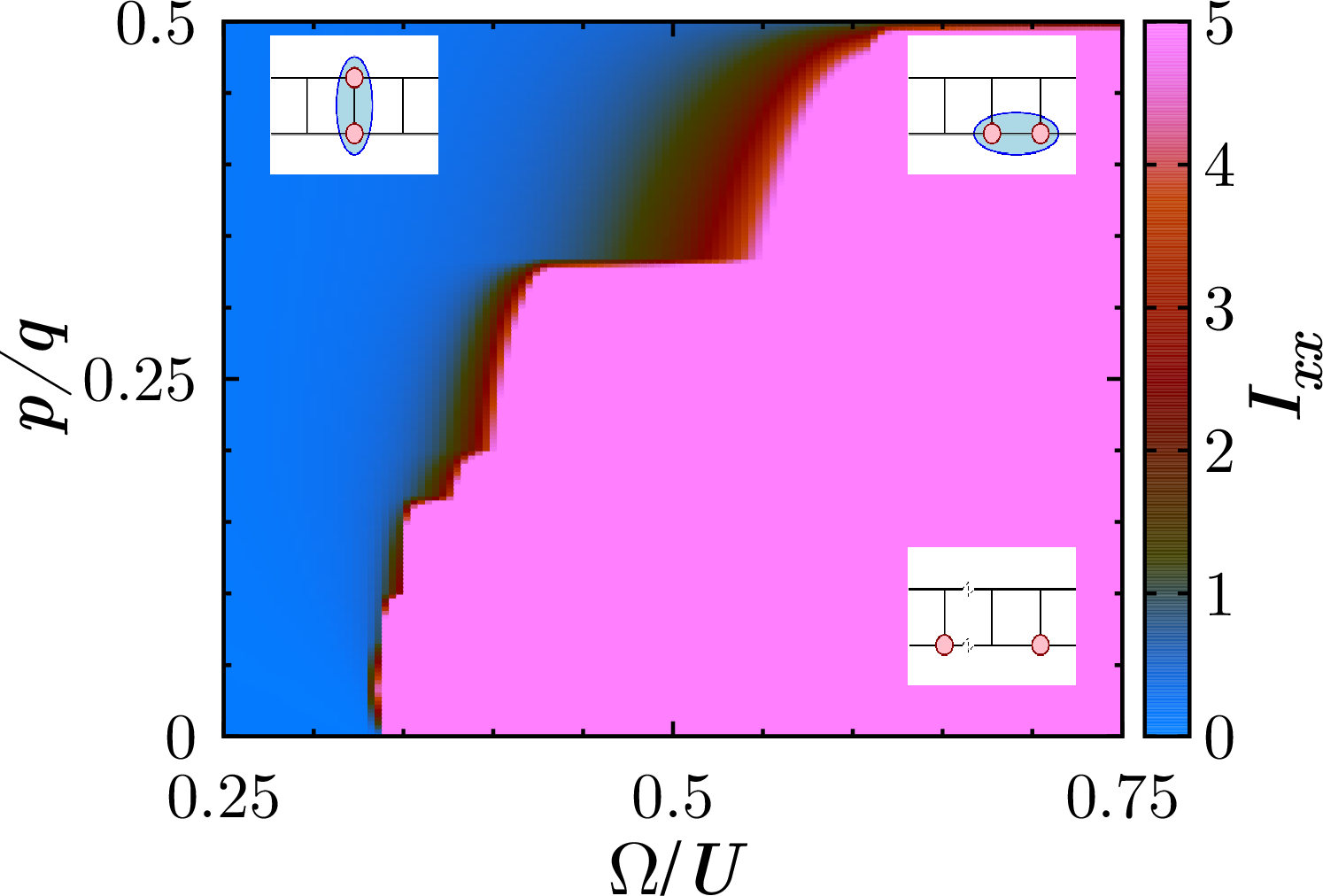}~~~~\includegraphics[height=0.8\myfigwidth,width=1.15\myfigwidth]{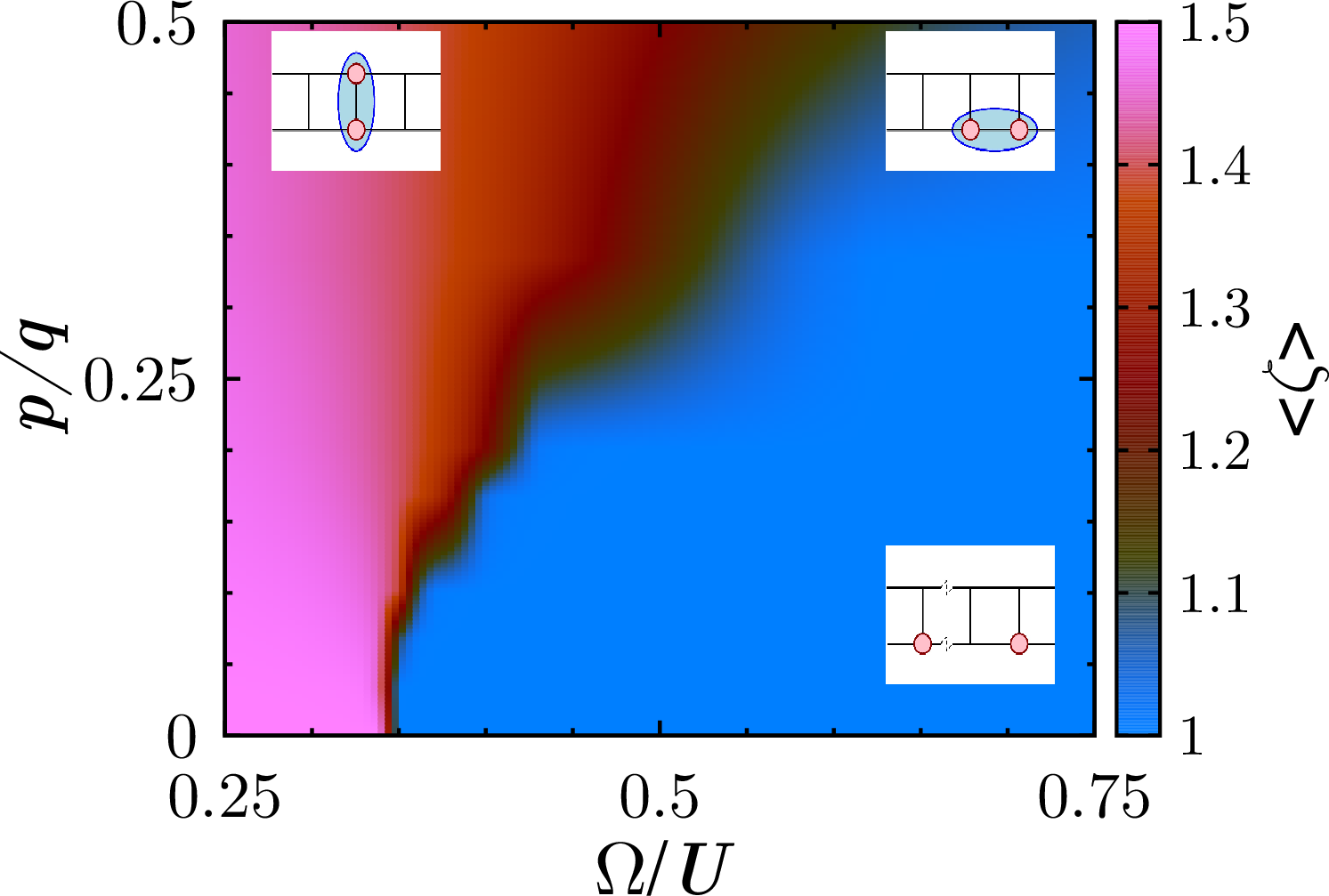}
{(a)~~~~~~~~~~~~~~~~~~~~~~~~~~~~~~~~~~~~~~~~~~~~~~~~~~~~~~~~~~~~~~~~~~~~~~~~~~~~~~~~(b)}

\centering
\includegraphics[height=0.5\myfigwidth,width=0.8\myfigwidth,trim= 0 12 0 0,clip]{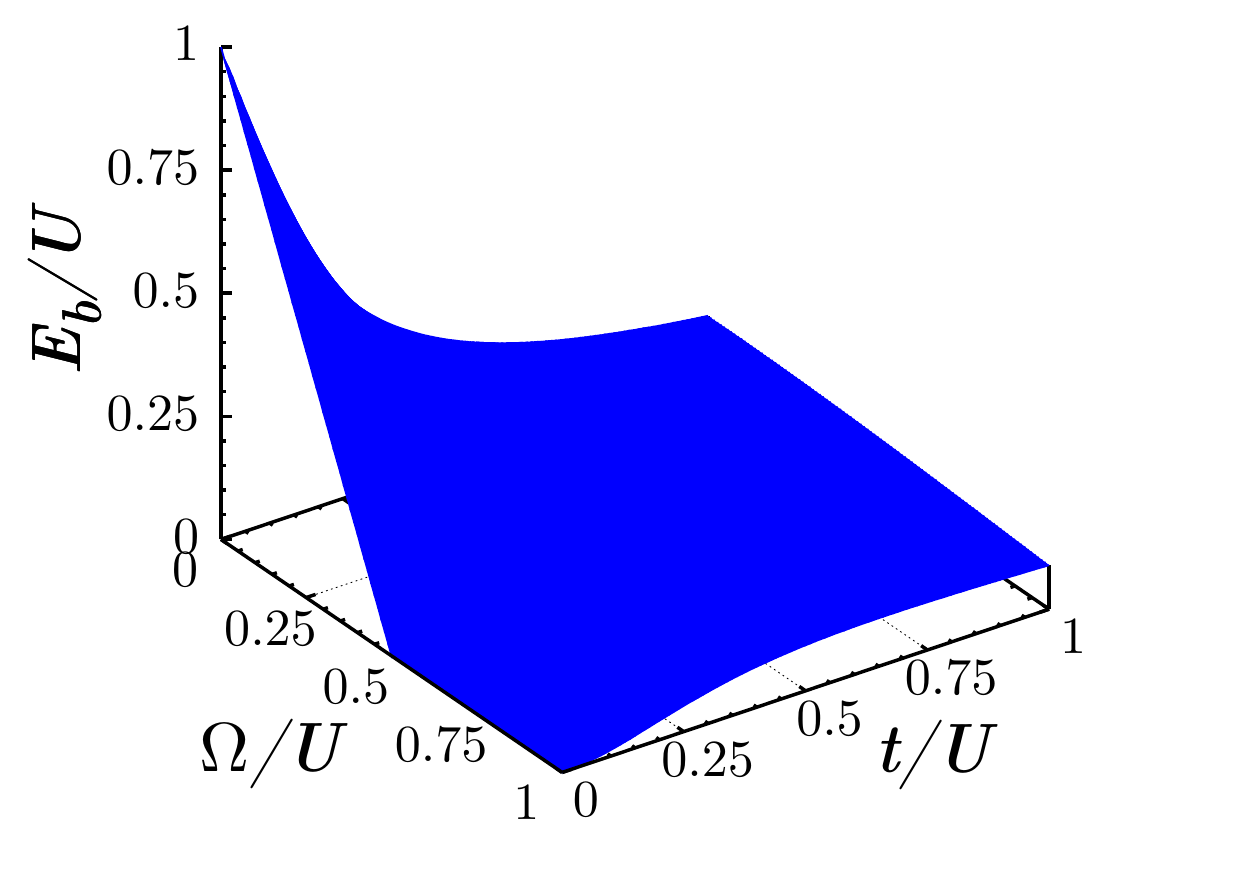}~\includegraphics[height=0.5\myfigwidth,width=0.8\myfigwidth]{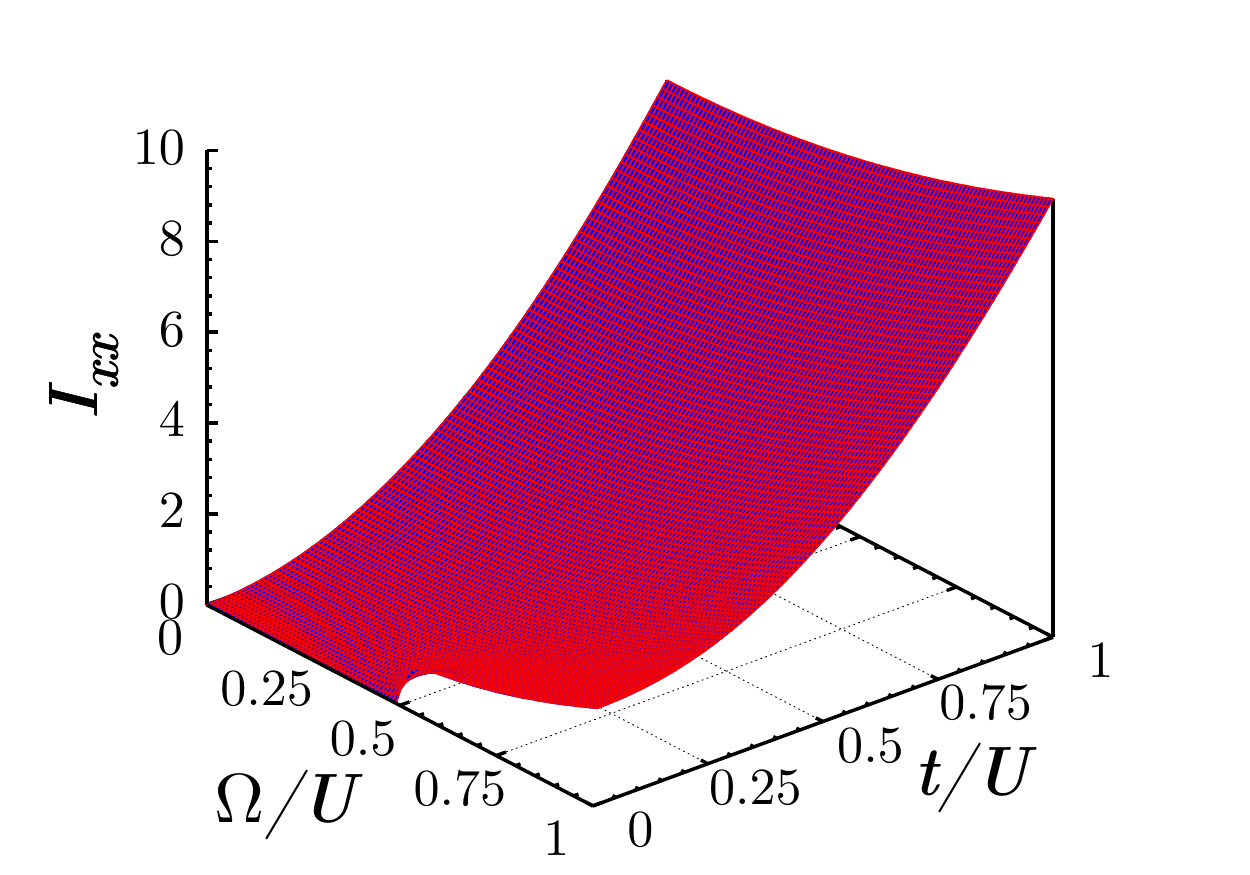}~\includegraphics[height=0.5\myfigwidth,width=0.8\myfigwidth]{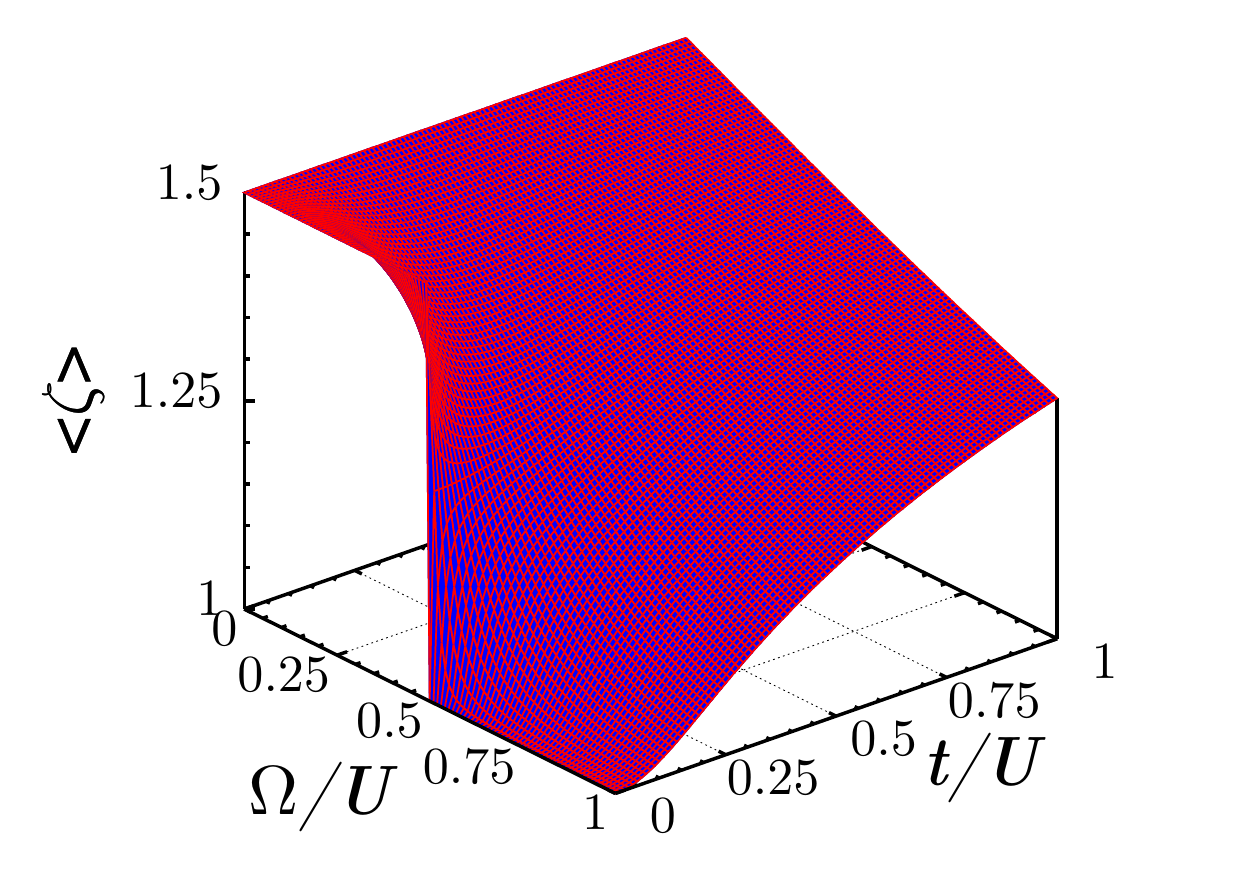}
{(c)~~~~~~~~~~~~~~~~~~~~~~~~~~~~~~~~~~~~~~~~~~~~~(d)~~~~~~~~~~~~~~~~~~~~~~~~~~~~~~~~~~~~~~~~~~~~~(e)}

\centering
\includegraphics[height=0.65\myfigwidth,width=1.05\myfigwidth]{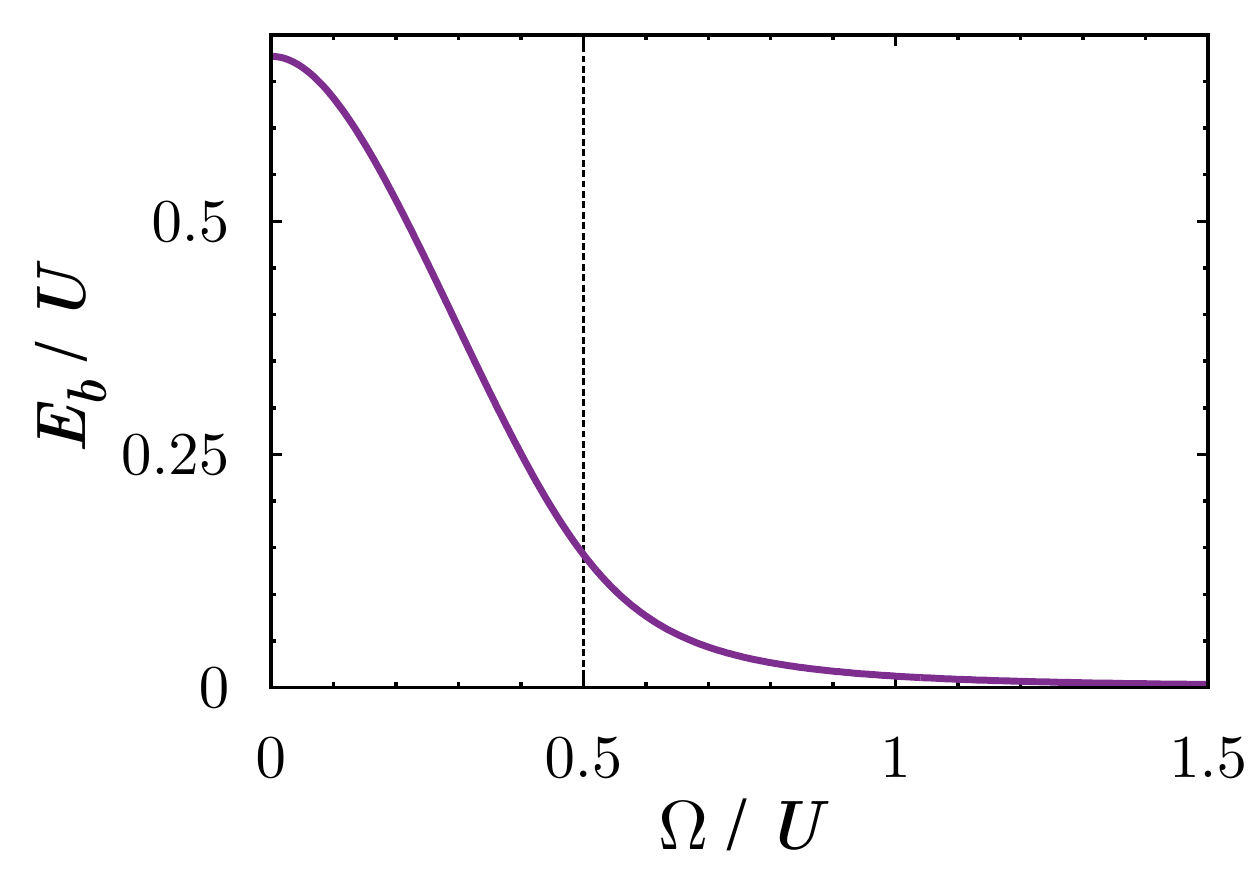}~~\includegraphics[height=0.65\myfigwidth,width=1.05\myfigwidth]{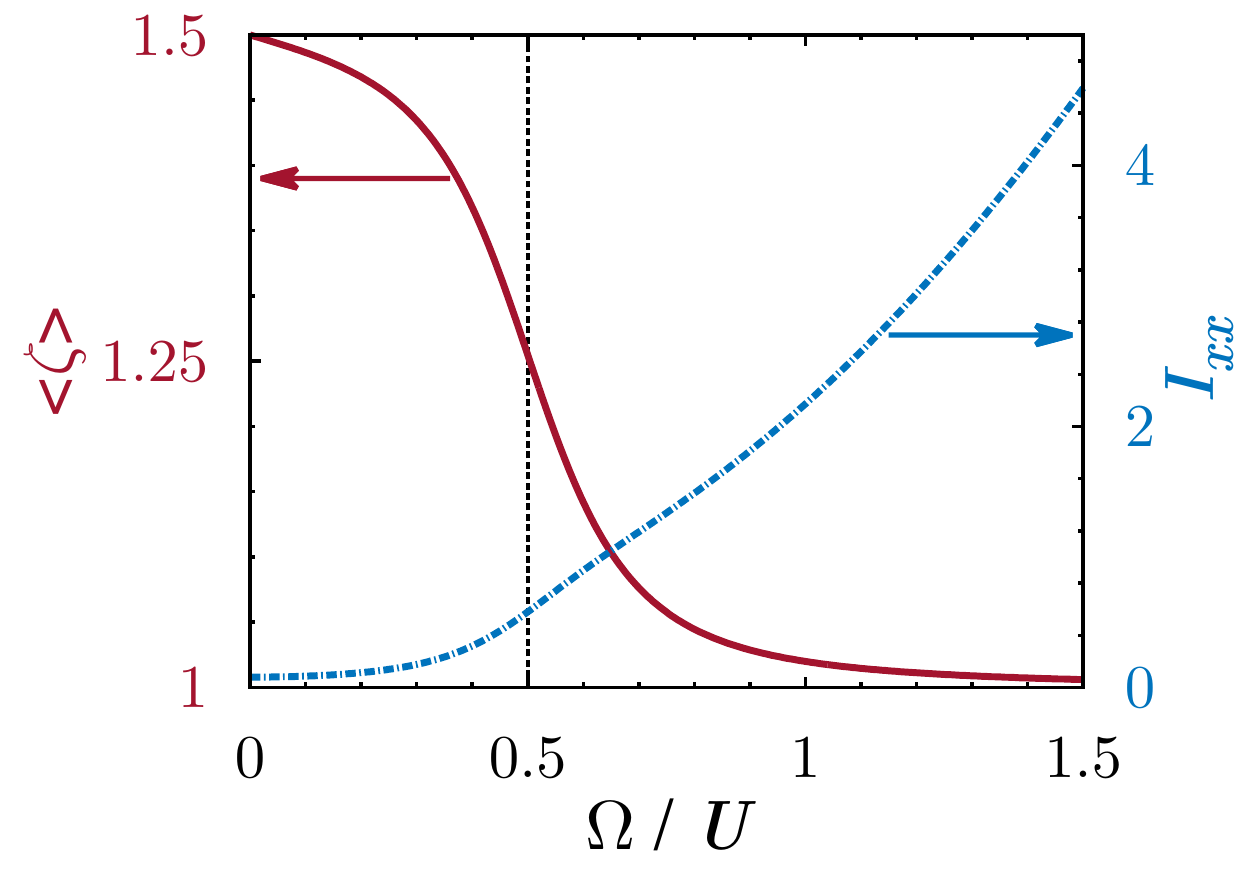}
{(f)~~~~~~~~~~~~~~~~~~~~~~~~~~~~~~~~~~~~~~~~~~~~~~~~~~~~~~~~~~~~~~~~~~~~~~~~~~~~~~~~~~~~~(g)}

\centering
\includegraphics[height=0.65\myfigwidth,width=1.05\myfigwidth]{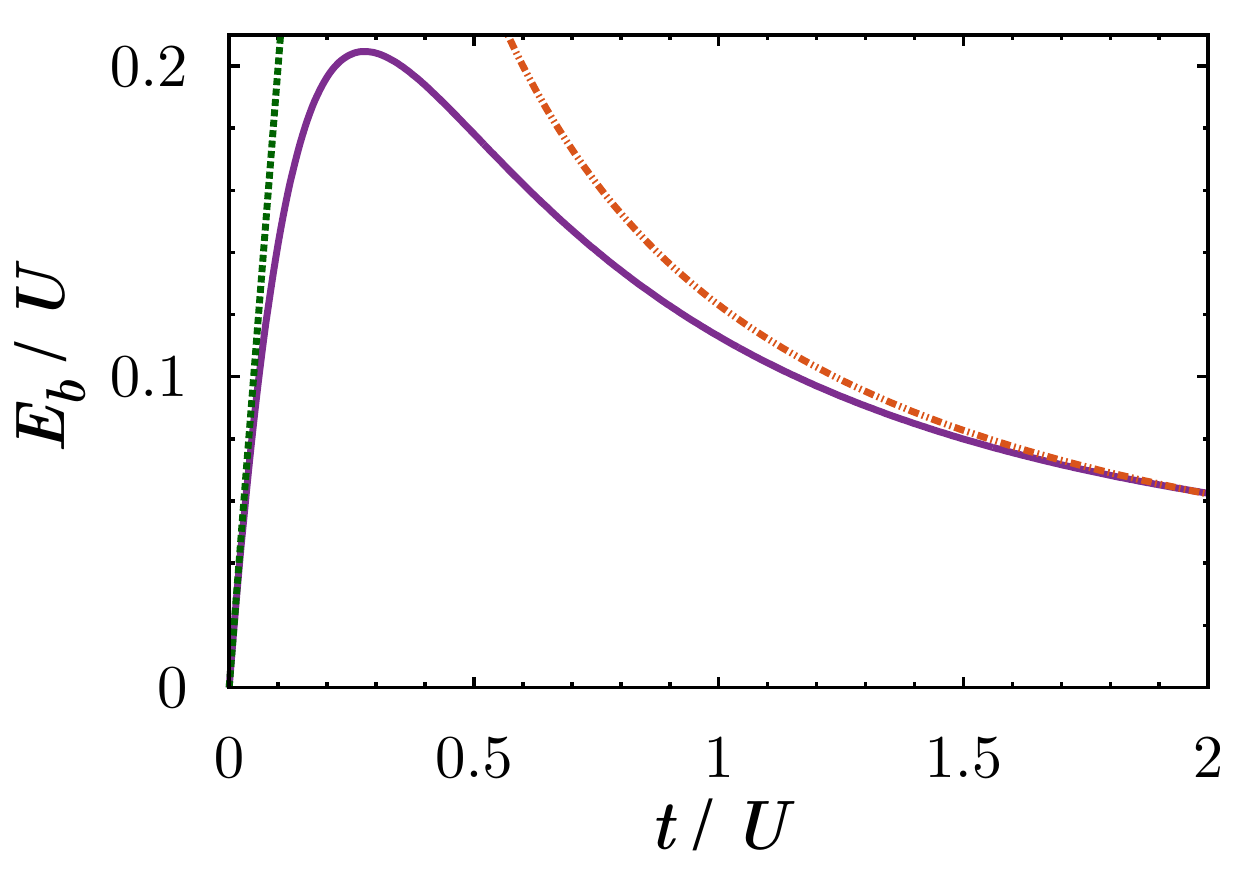}~~\includegraphics[height=0.65\myfigwidth,width=1.05\myfigwidth]{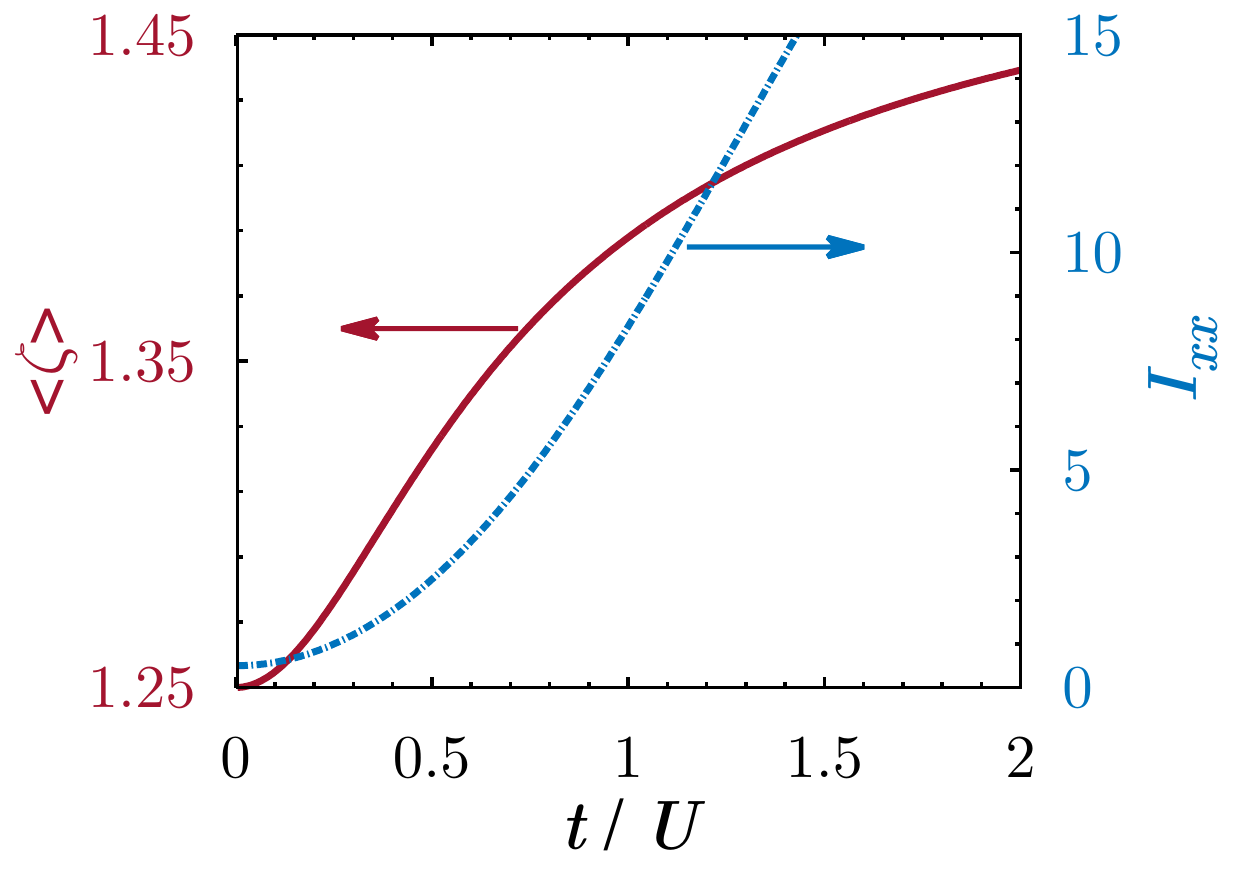}
{(h)~~~~~~~~~~~~~~~~~~~~~~~~~~~~~~~~~~~~~~~~~~~~~~~~~~~~~~~~~~~~~~~~~~~~~~~~~~~~~~~~~~~~~(i)}
}

\caption{(Color online) {\bf $M=2$ :} {\bf (a) and (b)} ``Phase diagram'' of two particles showing dependence (a) $I_{xx}$, and (b) $\mean{\zeta}$ on flux $p/q$ and $\Omega/U$ for $t/U=0.1$. Insets show the type of bound state stabilized.  {\bf (c) to (i) $\half$-flux} $E_b$, $I_{xx}$ and $\mean{\zeta}$ are respectively shown in (c), (d) and (e). (f) shows the dependence of $E_b$ on $\Omega/U$ for $t/U=0.1$, while (g) shows $I_{xx}$ and $\mean{\zeta}$ for the same case. Dependence of $E_b$ (h) and $I_{xx},\mean{\zeta}$ on $t/U$ at $\Omega = \Omega_c=U/2$. The dashed lines in (h) are results of analytical considerations at small and large $\frac{t}{U}$. 
} 
\mylabel{fig:M2}
\end{figure*}

\begin{figure*}
{
\centering
\includegraphics[height=0.8\myfigwidth,width=1.15\myfigwidth]{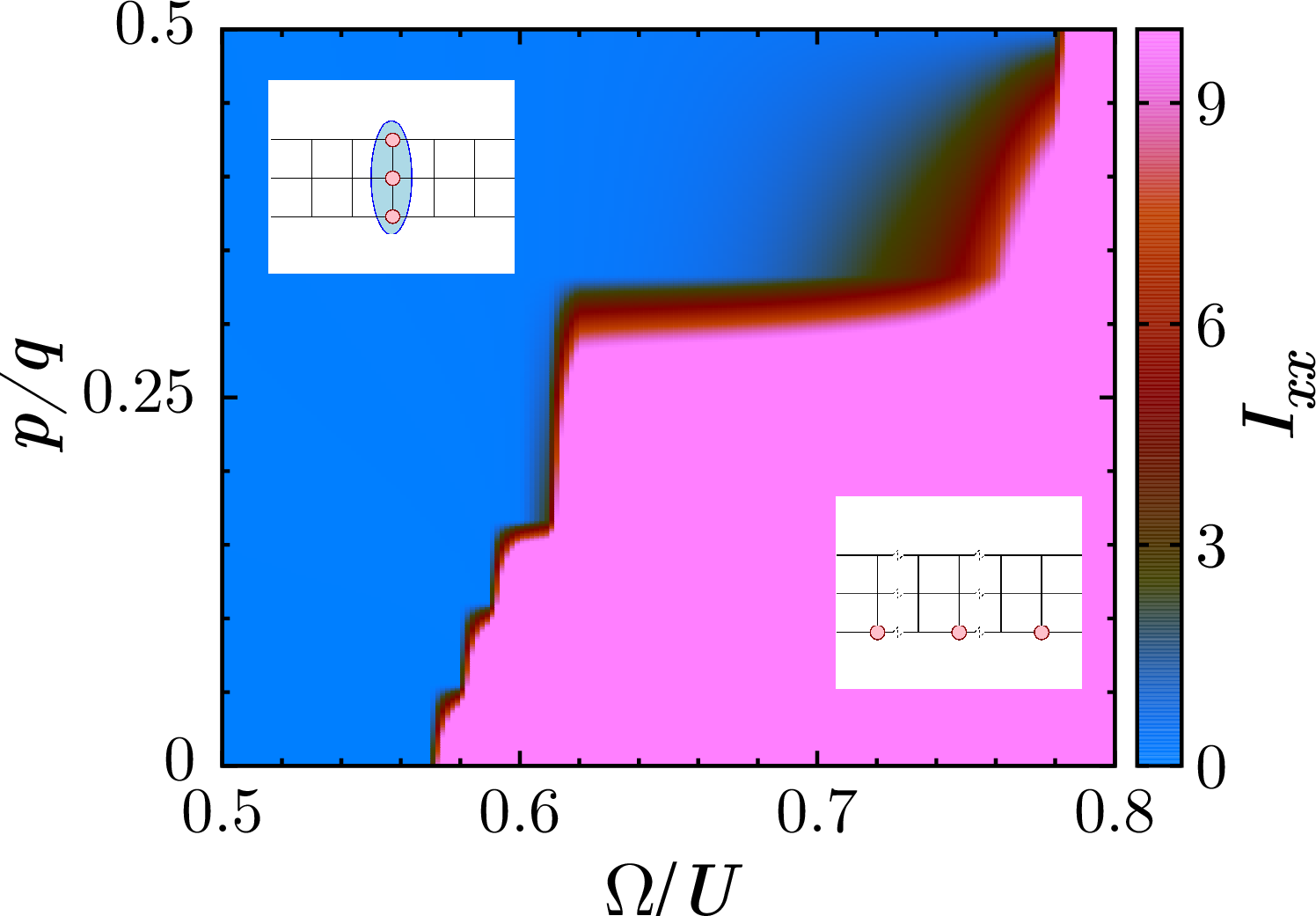}~~~~\includegraphics[height=0.8\myfigwidth,width=1.15\myfigwidth]{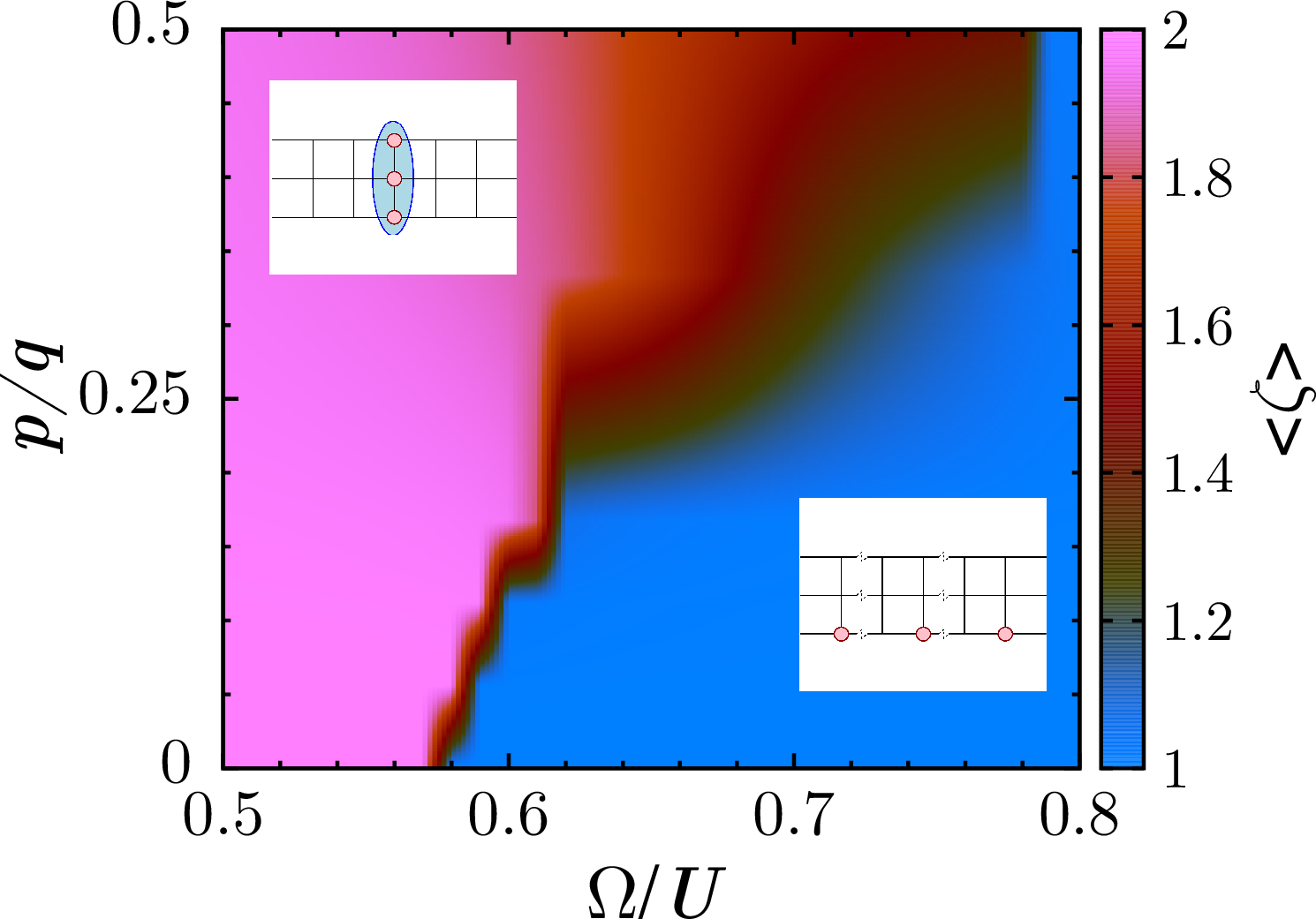}
{(a)~~~~~~~~~~~~~~~~~~~~~~~~~~~~~~~~~~~~~~~~~~~~~~~~~~~~~~~~~~~~~~~~~~~~~~~~~~~~~~~~~~~~~~~~~~~(b)}

\centering
\includegraphics[height=0.65\myfigwidth,width=1.05\myfigwidth]
{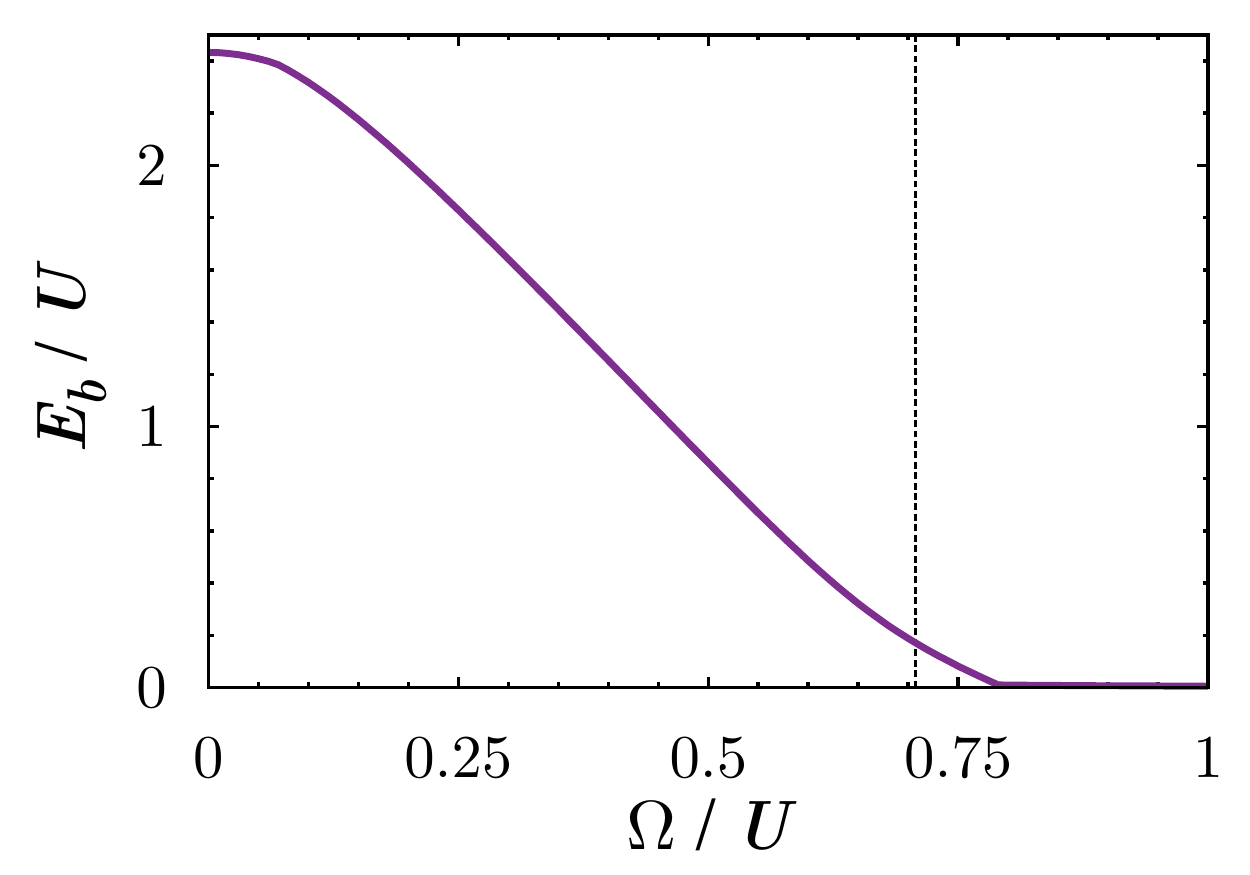}~~~~\includegraphics[height=0.65\myfigwidth,width=1.05\myfigwidth]{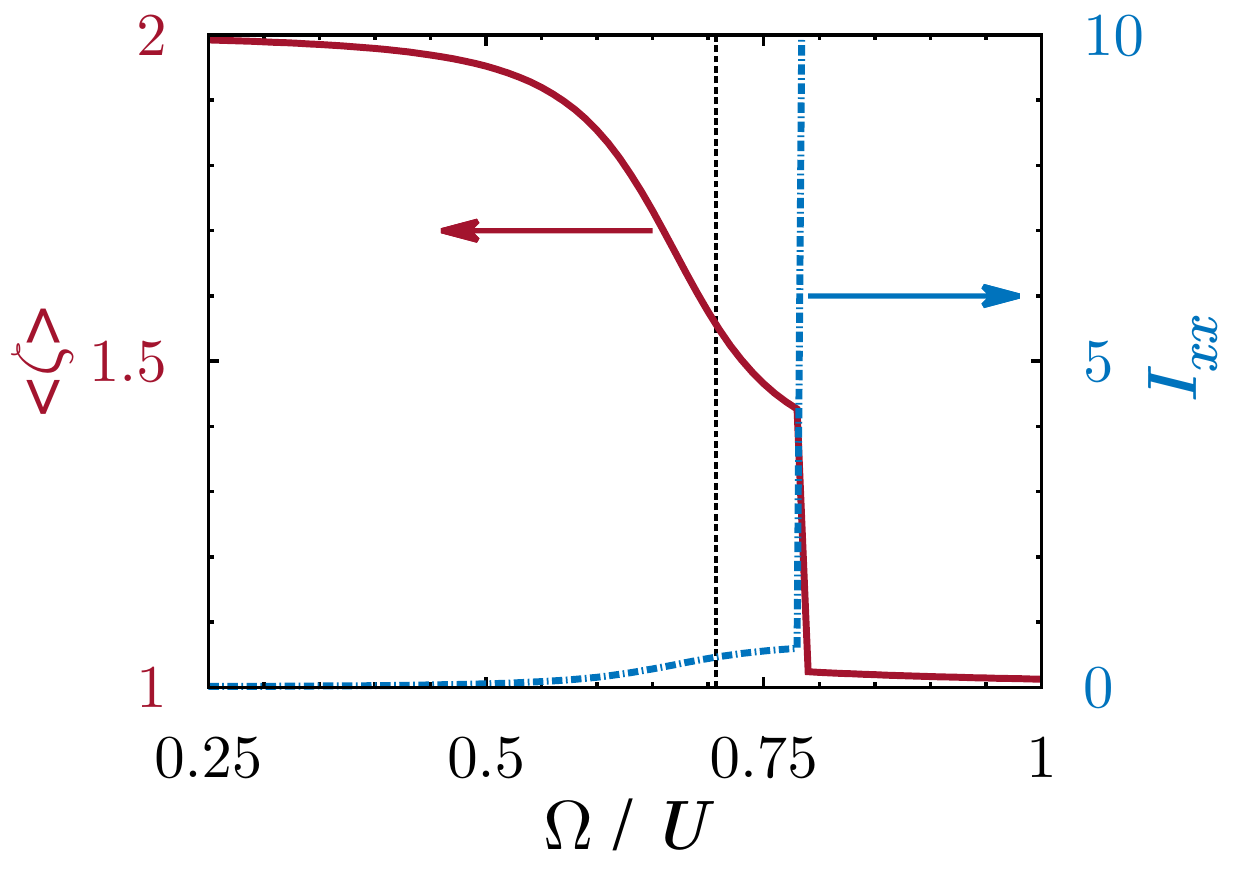}
{(c)~~~~~~~~~~~~~~~~~~~~~~~~~~~~~~~~~~~~~~~~~~~~~~~~~~~~~~~~~~~~~~~~~~~~~~~~~~~~~~~~~~~~~~~~~~~(d)}


\centering
\includegraphics[height=0.65\myfigwidth,width=1.05\myfigwidth]{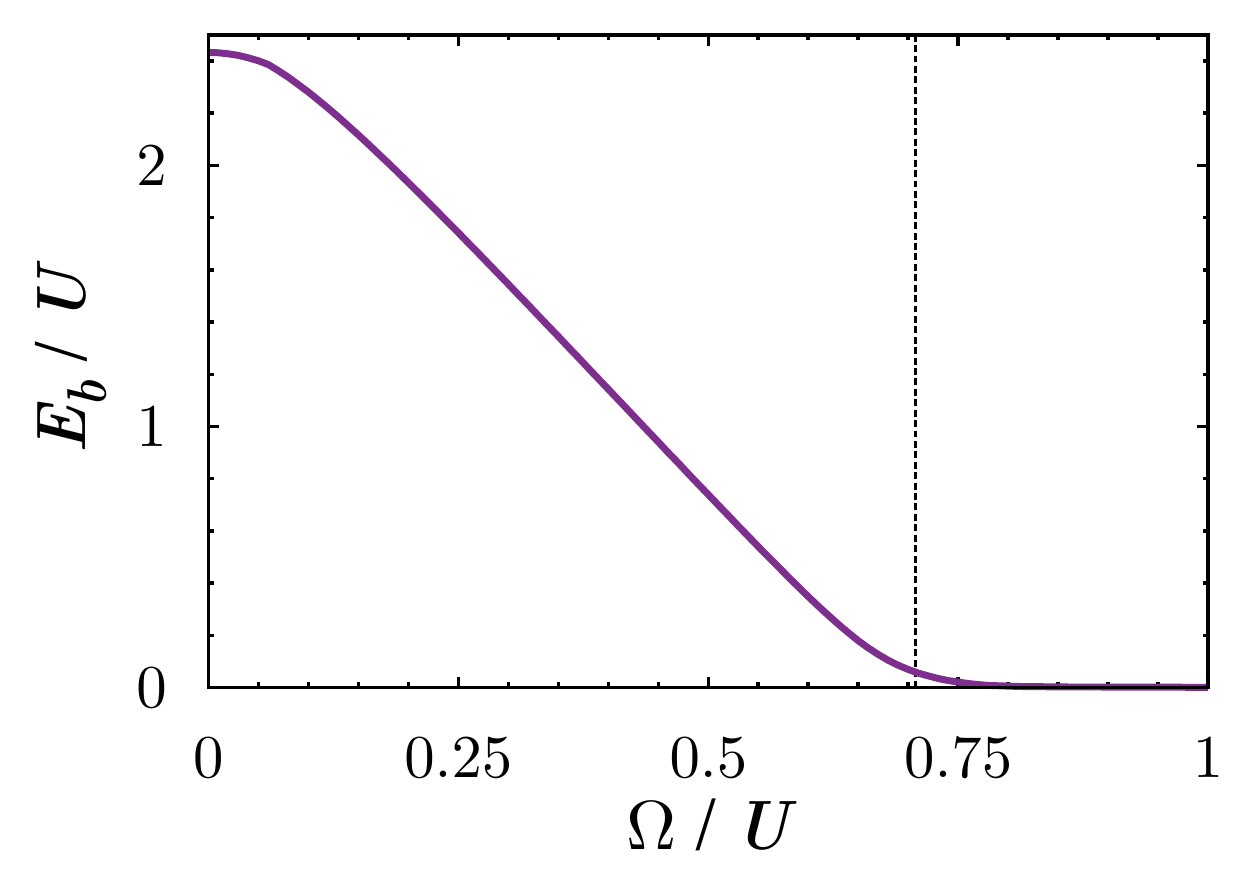}~~~~\includegraphics[height=0.65\myfigwidth,width=1.05\myfigwidth]{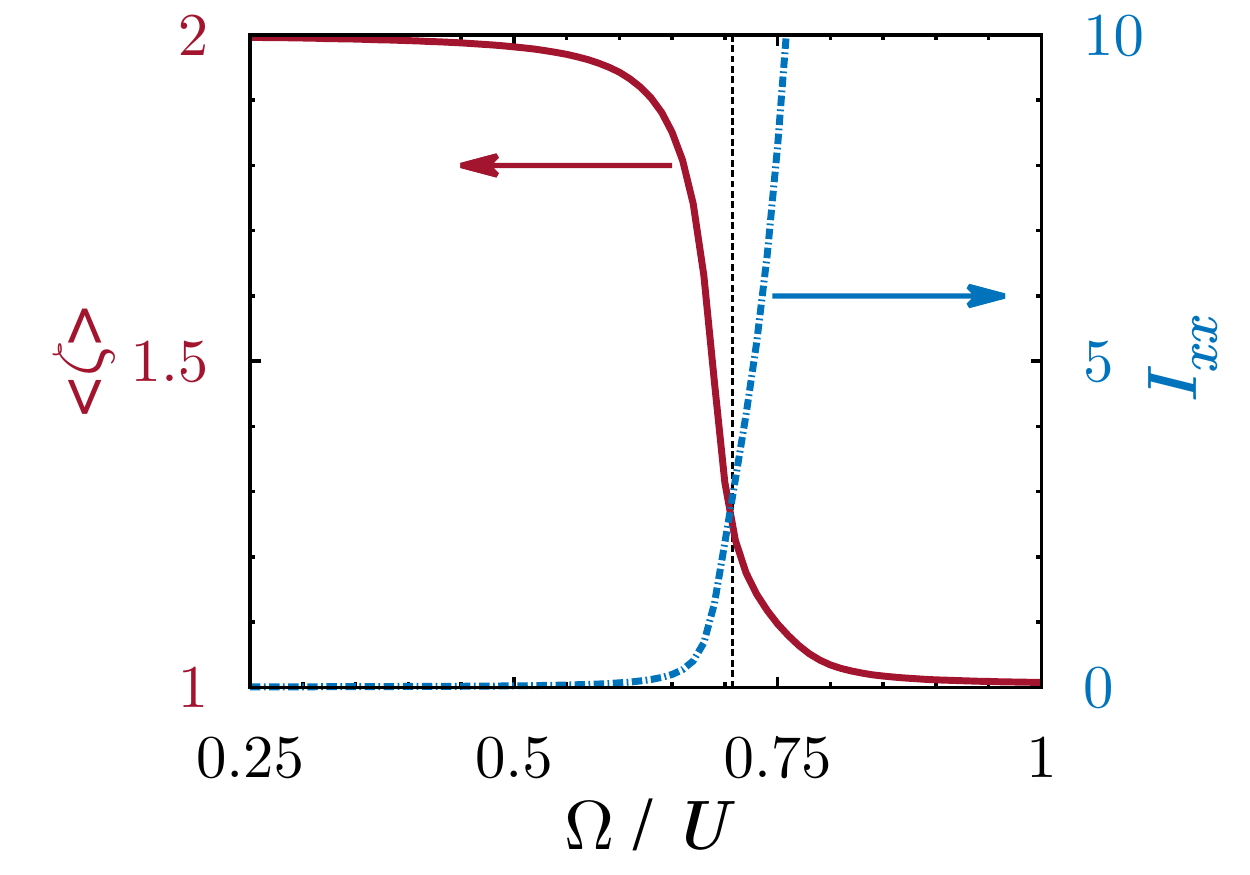}
{(e)~~~~~~~~~~~~~~~~~~~~~~~~~~~~~~~~~~~~~~~~~~~~~~~~~~~~~~~~~~~~~~~~~~~~~~~~~~~~~~~~(f)}
\centering
\includegraphics[height=0.65\myfigwidth,width=1.05\myfigwidth]{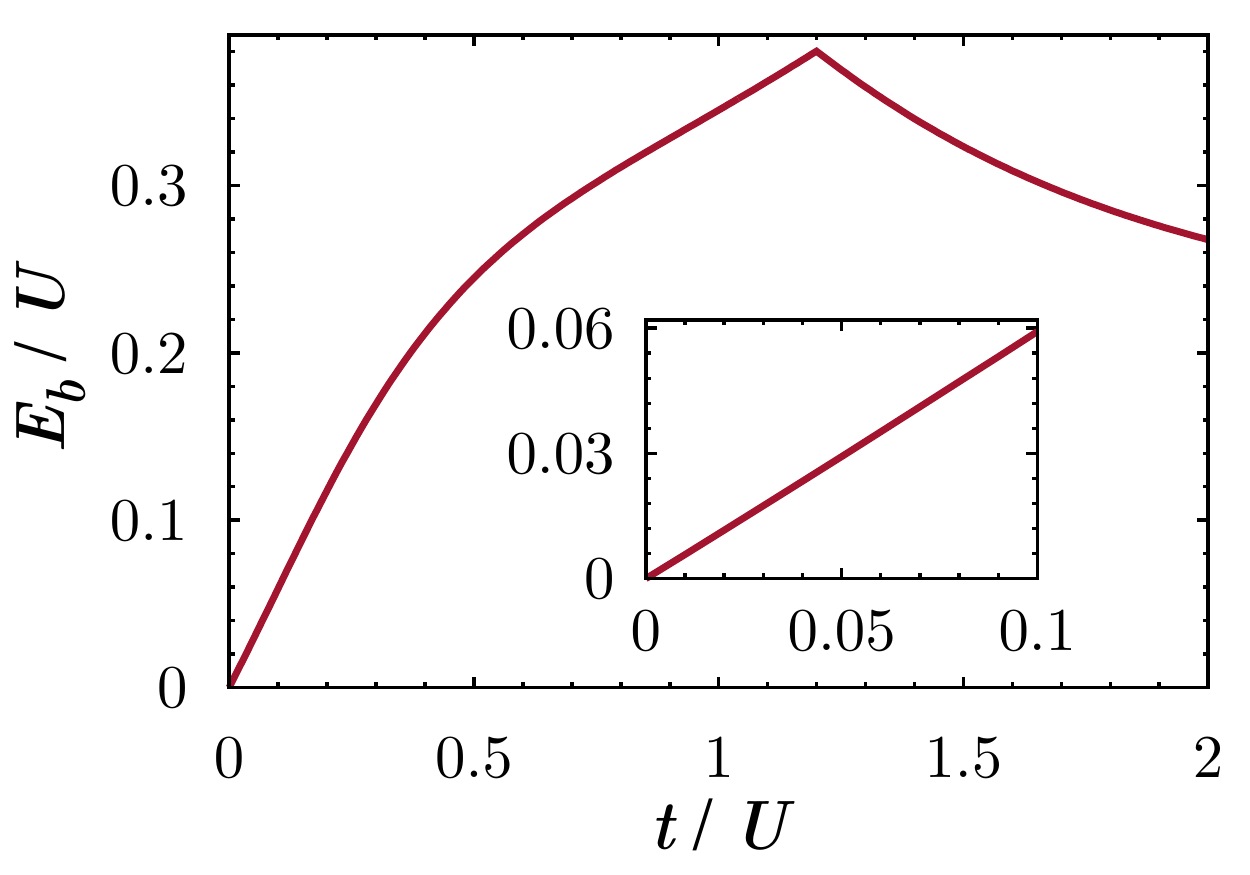}~~~~\includegraphics[height=0.65\myfigwidth,width=1.05\myfigwidth]{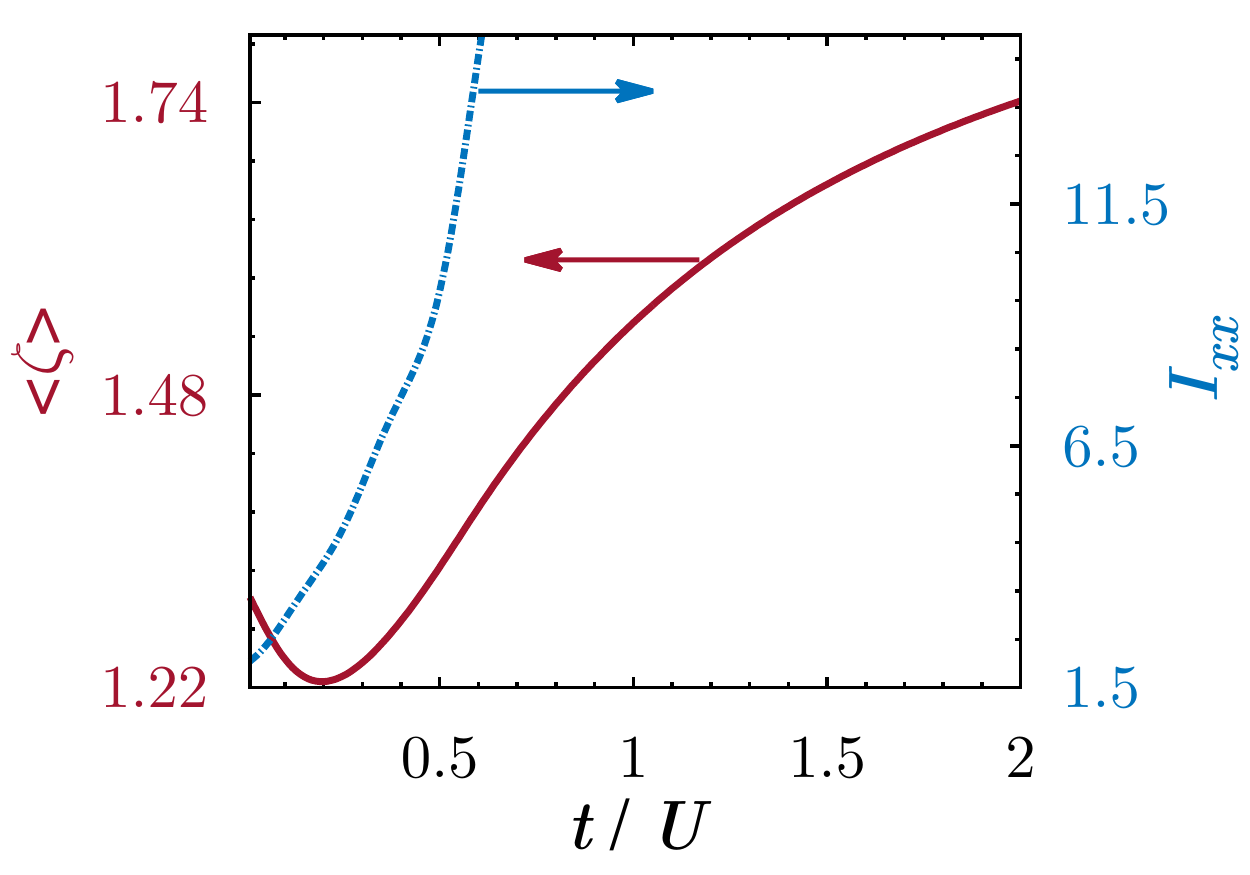}
{(g)~~~~~~~~~~~~~~~~~~~~~~~~~~~~~~~~~~~~~~~~~~~~~~~~~~~~~~~~~~~~~~~~~~~~~~~~~~~~~~~~~~~~~~~~~~~(h)}
}
\caption{(Color online) {\bf $M = 3$ :} {\bf (a) and (b) Phase diagram} 3-particle phase diagram showing dependence (a) $I_{xx}$, and (b) $\mean{\zeta}$ on flux $p/q$ and $\Omega/U$ for $t/U=0.1$ obtained with $N_q=18$. Insets show the type of bound state stabilized. {\bf (c) and (d), $\half$-flux :} Panels (c) and (d) show the dependence of the binding energy, and $I_{xx}$ and $\mean{\zeta}$ on $\Omega/U$ with $t/U=0.1$. {\bf (e) to (h), $p/q=1/3$:} (e) and (f) show same quantities as (c) and (d) for the $1/3$-flux case. (g) and (h) show the variation of $E_b$ and $I_{xx},\mean{\zeta}$ as a function of $t/U$ for $\Omega = \Omega_c = \frac{U}{\sqrt{2}}$ for the $1/3$-flux case.}
\mylabel{fig:M3}
\end{figure*}

\begin{figure*}
{
\centering
\includegraphics[height=0.8\myfigwidth,width=1.15\myfigwidth]{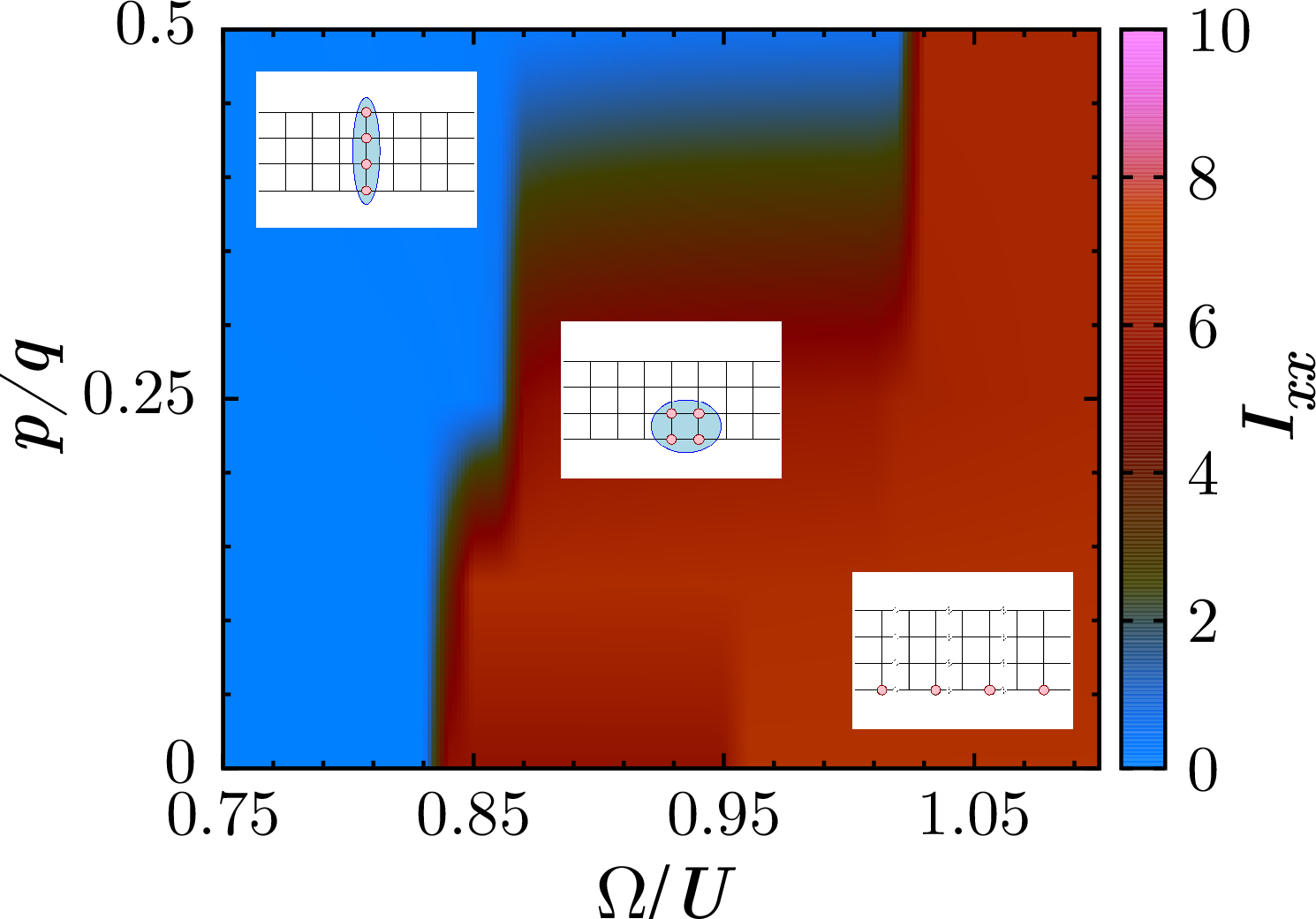}~~~~\includegraphics[height=0.8\myfigwidth,width=1.15\myfigwidth]{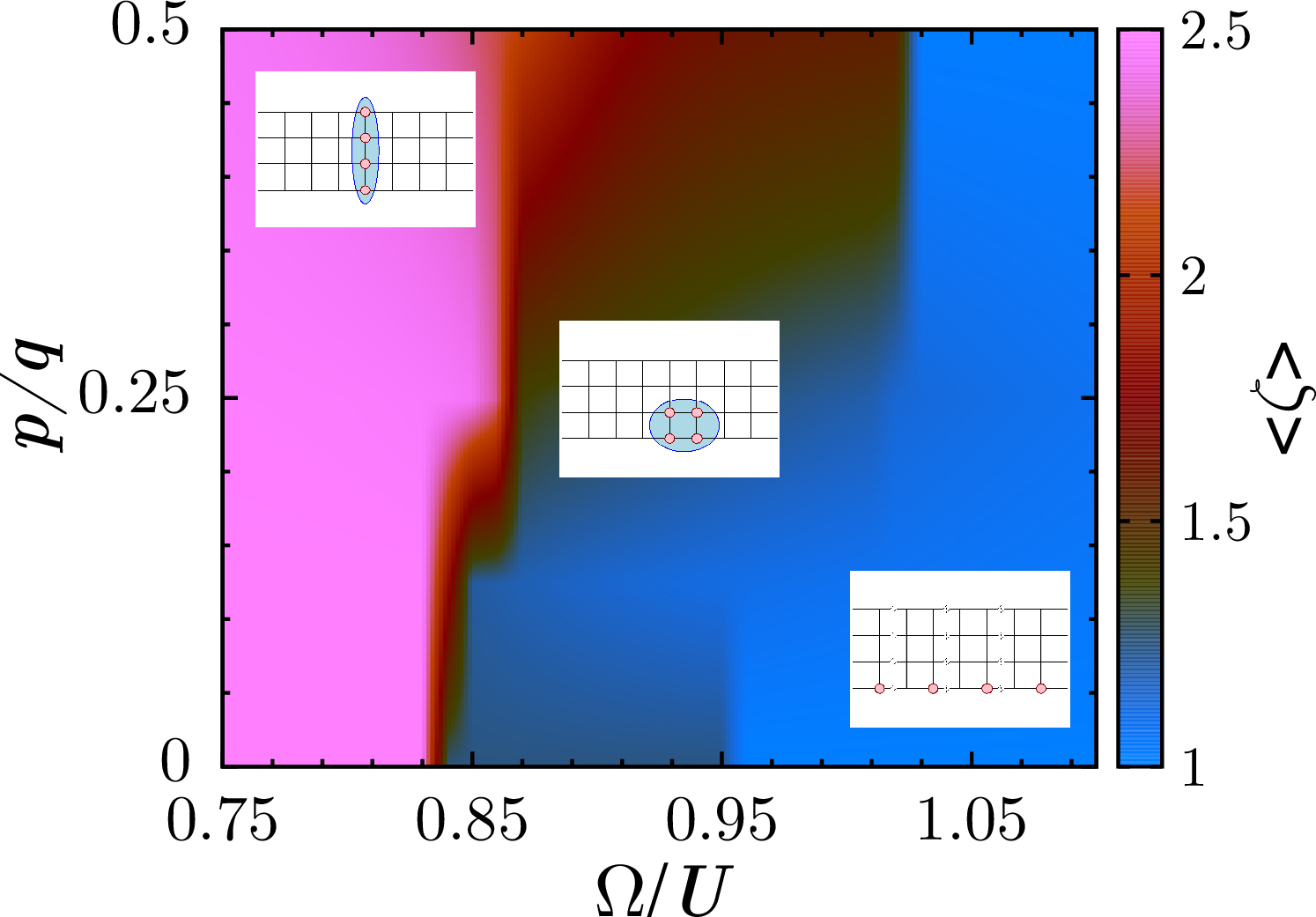}
{(a)~~~~~~~~~~~~~~~~~~~~~~~~~~~~~~~~~~~~~~~~~~~~~~~~~~~~~~~~~~~~~~~~~~~~~~~~~~~~~~~~(b)}

\centering
\includegraphics[height=0.65\myfigwidth,width=1.05\myfigwidth]{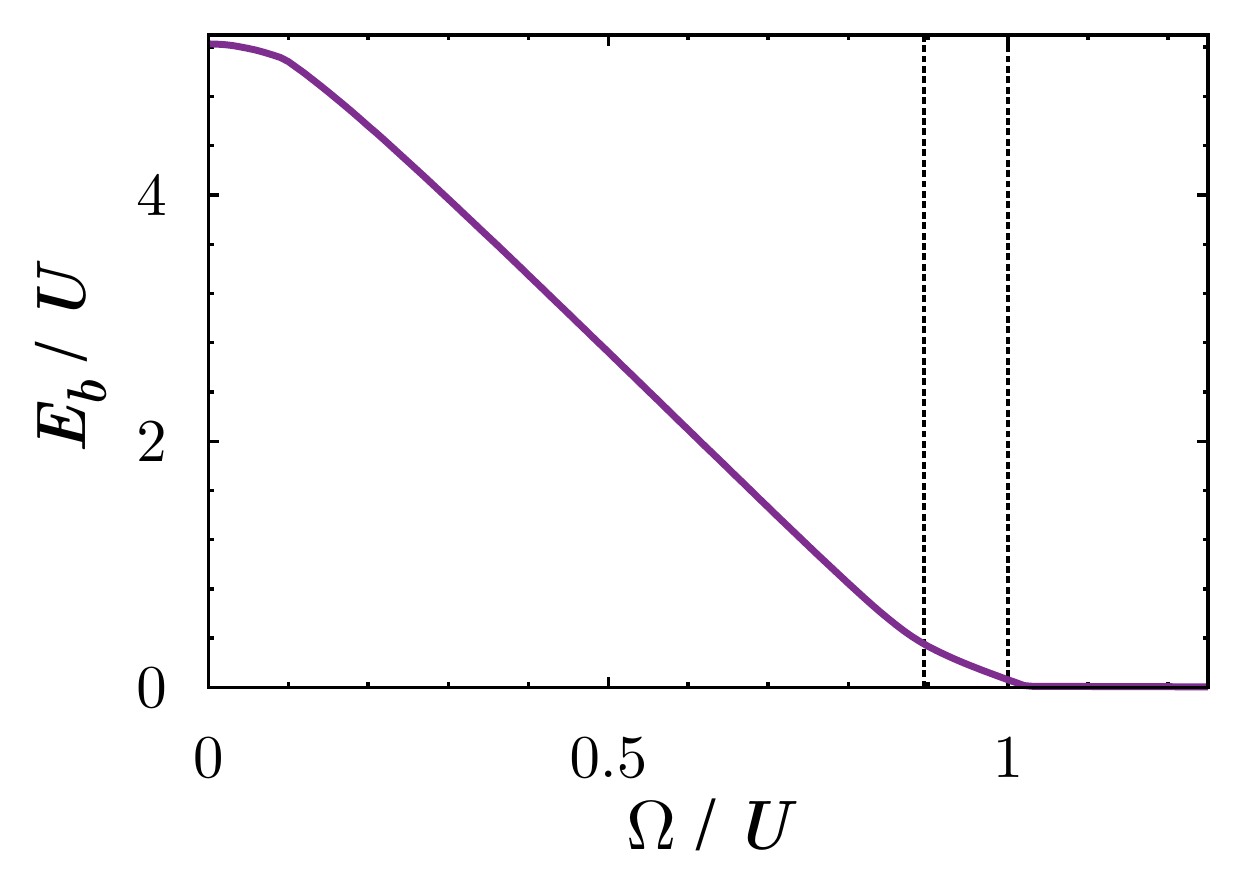}~~~~\includegraphics[height=0.65\myfigwidth,width=1.05\myfigwidth]{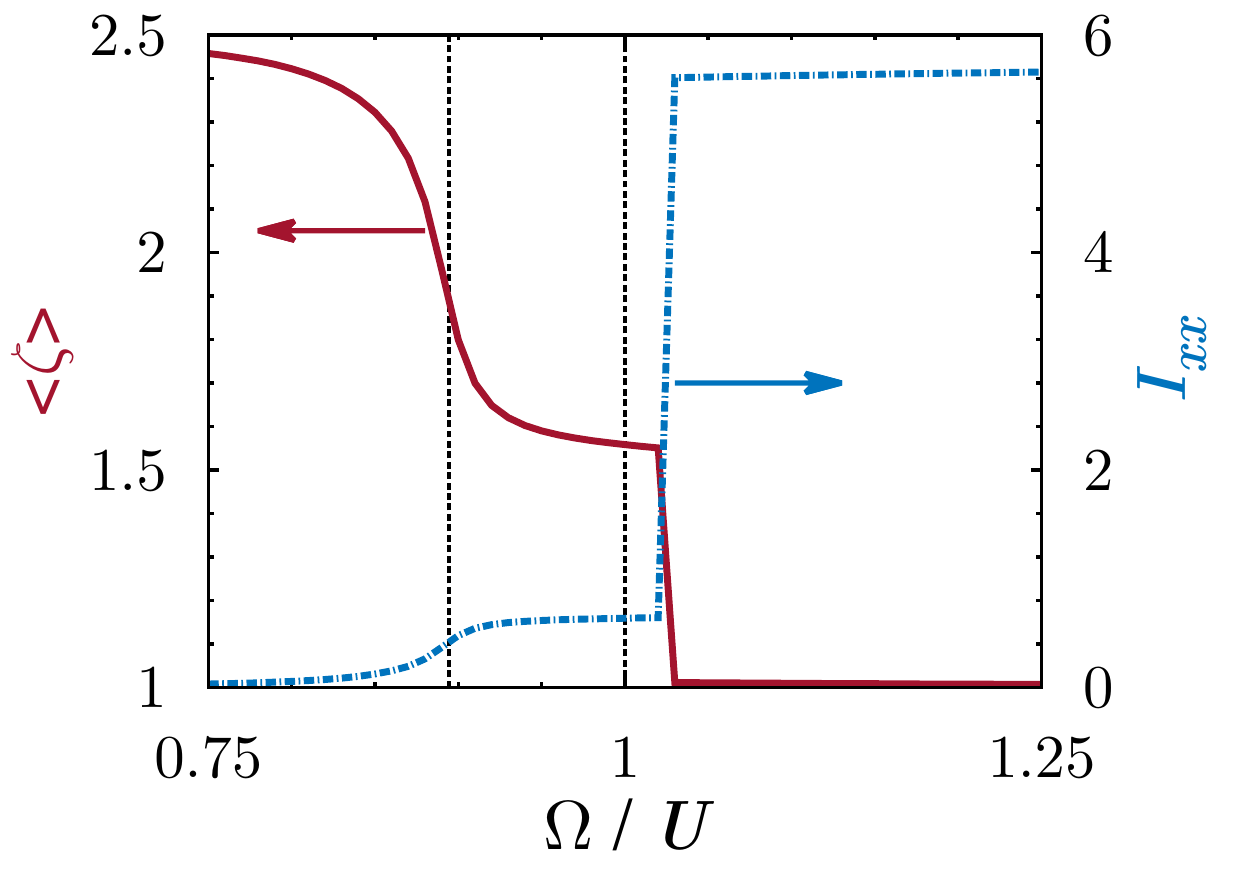}
{(c)~~~~~~~~~~~~~~~~~~~~~~~~~~~~~~~~~~~~~~~~~~~~~~~~~~~~~~~~~~~~~~~~~~~~~~~~~~~~~~~~~~~~~~~~~~~(d)}


\centering
\includegraphics[height=0.65\myfigwidth,width=1.05\myfigwidth]{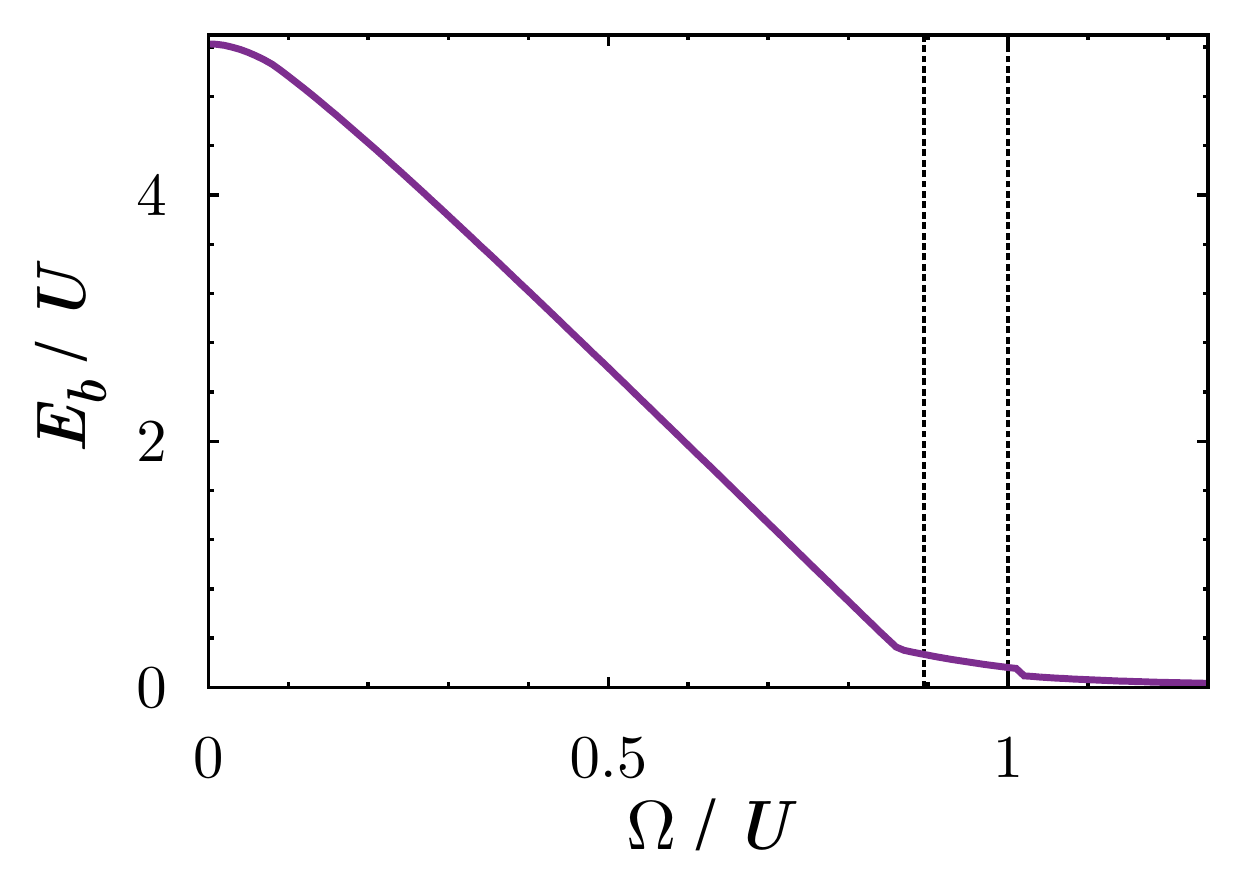}~~~~\includegraphics[height=0.65\myfigwidth,width=1.05\myfigwidth]{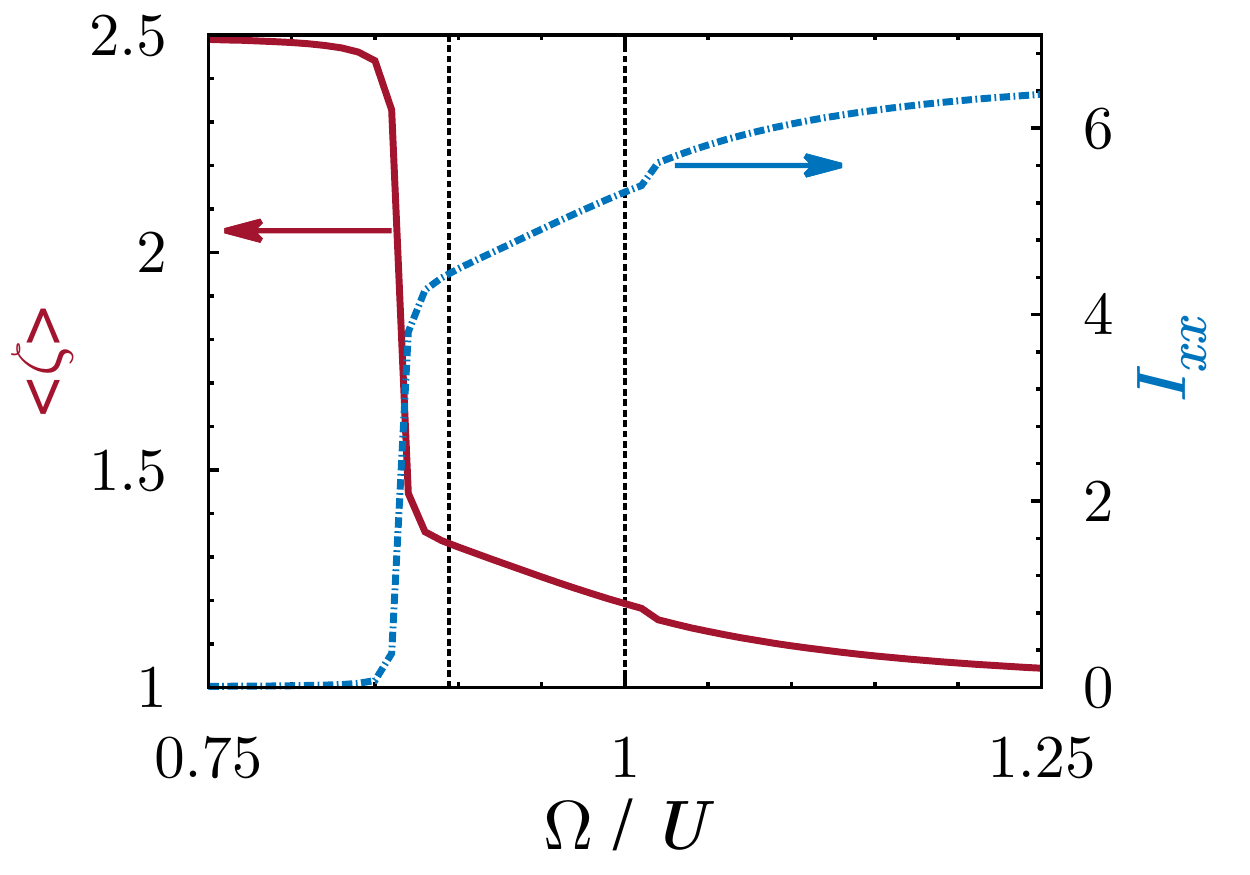}
{(e)~~~~~~~~~~~~~~~~~~~~~~~~~~~~~~~~~~~~~~~~~~~~~~~~~~~~~~~~~~~~~~~~~~~~~~~~~~~~~~~~~~~~~~~~~~~~(f)}

}
\caption{(Color online) (Color online) {\bf $M = 4$ :} {\bf (a) and (b) Phase diagram} 4-particle phase diagram showing dependence (a) $I_{xx}$, and (b) $\mean{\zeta}$ on flux $p/q$ and $\Omega/U$ for $t/U=0.1$ obtained with $N_q=8$. Insets show the type of bound state stabilized. {\bf (c) and (d), $\half$-flux:} Panels (c) and (d) show the dependence of the binding energy, and $I_{xx}$ and $\mean{\zeta}$ on $\Omega/U$ with $t/U=0.1$  {\bf (e) and (f), $p/q=1/4$:} (e) and (f) show same quantities as (c) and (d) for the $1/4$-flux case. }
\mylabel{fig:M4}
\end{figure*}

\noindent
{\bf Results:} While we choose the simplest case $\Omega_\gamma = \Omega$ to illustrate the physical ideas, our calculations can be readily adapted for specific experimental systems. \Fig{fig:M2} shows the results for $M=2$. In the absence of a flux $p/q\to 0$, the critical Zeeman field to break the baryon is  $\Omega_c = \half\left(\sqrt{U^2 + 16 t^2} - 4t \right)$. The phase diagram in the $p/q-\Omega$ plane shown in \fig{fig:M2}(a) and (b), show that this indeed happens at $p/q=0$. For larger $\Omega$, there is no bound state at $p/q=0$. Notice, however, that for $p/q=1/2$ ($\half$-flux), the situation is entirely different. One sees that the $I_{xx}$ remains finite with the increase in $\Omega$, and in fact $\mean{\zeta}$ goes to unity. As shown in the insets of the phase diagram the baryon evolves to the squished baryon.
We now investigate the $\half$ flux case which has this interesting physics in greater depth. Figs.~\ref{fig:M2}(c),(d) and (e) clearly demonstrate that for the $\half$-flux a bound state always exists (except when $t=0$) {\em irrespective} of a large Zeeman like field. This is a vivid example of the gauge field mitigating the baryon breaking effects of the Zeeman field. Fig.~\ref{fig:M2}(f) and (g) clearly demonstrate the squishing of the baryon by the Zeeman field, aided by the gauge field. Finally, \fig{fig:M2}(h) and (i) are to discuss the case where $\Omega=\Omega_c$. As shown from the analytic considerations (see \fig{fig:illus}), the binding energy of the squished baryon when $t \ll U$ is $\approx2t$ as it involves the hybridization process discussed in \fig{fig:illus}. Indeed, this result also corroborates quantitatively with the earlier arguments. For example, the binding energy at small $t$ is indeed found to be $2t$.

We now begin the discussion of the $M=3$ case, whose results are shown in \fig{fig:M3}, with a discussion of $\Omega_c$. When $t=0$, $\Omega_c = \frac{U}{\sqrt{2}}$ with a peculiar feature. Three distinct states are degenerate at this value of $\Omega$. There are the usual $M=3$ baryon\cite{Pohlmann13}, a completely broken baryon with three particles at different sites (``1+1+1''), and partially broken ``2+1'' baryon which has two particles at a given site with $\zeta=1$ and $2$ and the third particle at a different site with $\zeta=1$. \Fig{fig:M3}(a) and (b) show the phase diagram for this case in the $p/q$-$\Omega/U$ plane. Again the squishing effect of the Zeeman field aided by the non-Abelian gauge field is clearly seen. A crucial point is illustrated by the figs.~\ref{fig:M3}(c)-(f); the (c) and (d) panels of the figure are for the case with a $\half$-flux ($t/U=0.1$ ), which show the squishing of the baryon continuously (most rapidly near $\Omega_c$) with increase of $\Omega$. However, the process does not go on forever, and at a value of $\Omega$ somewhat larger than $\Omega_c$, the baryon completely breaks up and there is no bound state. Here, therefore, the gauge field produced by the $\half$-flux is unable to prevent the pair breaking effect. Most interestingly, the situation changes completely if one introduces a $1/3$-flux. As shown in \fig{fig:M3}(e) and (f), the squishing occurs smoothly, and in fact, we believe, that there is a bound state for all $\Omega$ (we cannot verify this as binding energy becomes small with a concomitant large $I_{xx}$). Another interesting result is that at $\Omega_c$, the binding energy for small $t$ can be inferred to be proportional to $t$. This is due to the hybridization between the ``2+1'' baryon hybridizing with a ``1+1+1'' aided by the $1/3$-gauge field (flavor-orbit coupling). Indeed, as shown in \fig{fig:M3}(g), this is in excellent agreement with the exact result.

Results for $M=4$ are shown in \fig{fig:M4}. The novel aspect here is the presence of two critical Zeeman fields $\Omega_{c1}$ and $\Omega_{c2}$. When $t=0$, the usual 4-baryon is destabilized to state with two $2$-baryons (each of which can be located at any site) at $\Omega_{c1} = \frac{2 U}{\sqrt{5}}$. At $\Omega_{c2} = U$ this state  is again broken into a $1+1+1+1$ state where each particle can be at any site distinct from others with $\zeta=1$. With a $\half$-flux, one sees a smooth transformation form the usual 4-baryon to a $2$+$2$ baryon (bound state of 2-baryons) -- a vivid example of squishing. However, the $\half$-flux is not able to mitigate the effects of the Zeeman field. In fact, near $\Omega_{c2}$ the squished $2$+$2$ baryon is broken up. Remarkably, for a flux of $1/4$, this transition is prevented, and our calculations suggest bound state for any $\Omega$ (checking this requires very large computational resources). At $\Omega_{c1}$, we can also show that the binding energy is proportional to $t^2$ (in-order to hybridize the 4-baryon and the 2+2 baryon); our exact calculations (not shown here) have borne this out. 

We now discuss the general criteria that are required to produce squishing (rather than breaking). Clearly, the non-abelian gauge field induced by the flux must be able to hybridize the two (or more) degenerate states that occur at the critical Zeeman fields. For example, the flavour-orbit coupling with a $\half$-flux does not hybridize the $2$+$2$ state with the $1+1+1+1$ state for $M = 4$, and hence the baryon is broken rather than squished. Knowing the details of $\Omega_\gamma$, one can choose appropriate flux to achieve this for a given system.

 In the many body setting, there is clearly a rich collection of phases and crossovers that can be explored. In particular, quasi-condensates of squished baryons are likely to hold interesting physics. The few body results developed here can be used as a guide for such studies particularly in the dilute limit.

We conclude the paper by pointing out an interesting possibility to use this system to create a class Hamiltonians call ``random flux'' models\cite{Altland1999}. The idea is to introduce some randomness in $\Omega_\gamma^j$, which in turn will make the gauge fields (\eqn{eqn:GF}) also random. If the lattice chosen is a square lattice, the Hamiltonian of the type (\eqn{eqn:GF}) realized will be similar to a ``random flux'' model.

\noindent
{\bf Acknowledgments:}  S.~K.~G. acknowledges  support from CSIR, India via  SRF grants. U.~K.~Y. acknowledges support from UGC, India via Dr. D.~S. Kothari Post-doctoral Fellowship scheme. V.~B.~S.~is grateful to DST, India and DAE, India (SRC grant) for generous support. We thank Sambuddha Sanyal for discussions regarding the random flux model, Adhip Agarwala and Aabhaas Mallik for comments on the manuscript.

\bibliography{refbsquish}

\end{document}